# MODELLING OF AUTOIGNITION AND NO SENSITIZATION FOR THE OXIDATION OF IC-ENGINE SURROGATE FUELS

J.M. ANDERLOHR[1,2*], R. BOUNACEUR[2], A. PIRES DA CRUZ[1], F. BATTIN-LECLERC[2]

[1]IFP,

1 et 4, Ave. Bois Préau,

92852 Rueil Malmaison Cedex, France.

[2]Département de Chimie-Physique des Réactions,

UMR n°7630 CNRS, INPL-ENSIC,

1 rue Grandville, BP 20451, 54001 NANCY Cedex, France.

Full-length article

SHORTENED RUNNING TITLE:

MODELLING OF AUTOIGNITION AND NO SENSITIZATION

[*] E-mail : Jorg.ANDERLOHR@ifp.fr; Tel.: +33 1 47 52 60 00

ABSTRACT

This paper presents an approach for modelling with one single kinetic mechanism the chemistry of the autoignition and combustion processes inside an internal combustion engine, as well as the chemical kinetics governing the post-oxidation of unburned hydrocarbons in engine exhaust gases. Therefore a new kinetic model was developed, valid over a wide range of temperatures including the negative temperature coefficient regime. The model simulates the autoignition and the oxidation of engine surrogate fuels composed of n-heptane, iso-octane and toluene, which are sensitized by the presence of nitric oxides. The new model was obtained from previously published mechanisms for the oxidation of alkanes and toluene where the coupling reactions describing interactions between hydrocarbons and $NO_x$ were added. The mechanism was validated against a wide range of experimental data obtained in jet-stirred reactors, rapid compression machines, shock tubes and homogenous charge compression ignition engines. Flow rate and sensitivity analysis were performed in order to explain the low temperature chemical kinetics, especially the impact of $NO_x$ on hydrocarbon oxidation.

**INTRODUCTION**

During the last decades, European Union legislation has imposed more and more stringent restrictions on vehicle exhaust emissions forcing current engine research and development to focus strongly on the reduction of pollutant emissions. The main pollutants emitted from gasoline Internal Combustion (IC) engines are compounds of condensed fuel, Unburned HydroCarbons (UHC) and Nitric Oxides ($NO_x$). Different strategies for reducing UHC and $NO_x$ emissions are known.

The formation of pollutants inside the engine can be minimised by optimising the internal combustion processes. This is the case for IC-engines controlled via Homogenous Charge Combustion Ignition (HCCI). HCCI engines combine advantages of a Spark Ignition (SI) engine with those of a Compression Ignition (CI) engine. The homogenous fuel/air mixture guarantees low particulate emissions and the high dilution permits very low production of $NO_x$, while the principle of CI assures a high efficiency close to that of a diesel engine [1-3]. HCCI engines operate at very lean conditions and are highly diluted via Exhaust Gas Recirculation (EGR). The principle of EGR is to re-inject an exhaust gas fraction into the combustion chamber together with the fresh gases. EGR has the advantage of reducing the maximum combustion temperature in the engine resulting in a reduced production rate of thermal-NO. However, HCCI exhaust gases are composed of many different species, including UHC and traces of $NO_x$. Thus, recycled UHC and $NO_x$ will interact inside the combustion chamber with the injected fuel and impact oxidation kinetics and ignition delays [4].

During compression in EGR-controlled HCCI engines the cool flame phenomenon may appear and control the fuel autoignition. A cool flame is the result of a limited exothermic reaction that is associated with a partial conversion of the fuel and appears during fuel oxidation at low temperatures (650–800 K). Recycled UHC, CO and $NO_x$ mixed by EGR with the fresh gases may strongly impact the fuel oxidation. In particular the impact of $NO_x$ on HC oxidation is known to be complex [4-7]. Predicting the impact of such interactions on the control of autoignition delays demands an understanding of the complex chemical kinetics at low temperatures.

Other strategies aim on minimizing pollutant emissions by their reduction in the exhaust line by catalyst oxidation and reduction. Secondary Air Injection (SAI) might be one such strategy. Operated at temperatures below the catalysts light-off limit (~300°C) the catalytic converter is mostly ineffective during engine cold start. There is a strong interest in rapidly achieving the catalysts thermal operating conditions [8]. The injection of air close to the exhaust valve induces a post-oxidation of the UHC in the exhaust-gas. The post combustion takes place in the exhaust pipe between the combustion chamber and the catalytic converter. The heat released due to the oxidation of the UHC heats up the catalytic converter and therefore reduces the time taken for the catalytic converter to reach its light-off temperature [9-11]. Thermochemical conditions of HC and UHC post-oxidation in the exhaust line are restrictive. The stoichiometry of exhaust gases may locally vary from rich at the cylinder exit to very lean after SAI. Gas temperatures may vary between 600 and 900 K [9], which is the zone where post-oxidation processes are governed by complex chemical kinetics and where a Negative Temperature Coefficient (NTC) behavior is observed.

The main purpose of this paper is to model the chemical kinetics of gasoline oxidation in presence of UHC and $NO_x$. Gasoline is mostly composed of saturated hydrocarbons (normal and branched alkanes) and aromatics. N-heptane and iso-octane are Primary Reference Fuels (PRF) for octane rating in IC-engines [12, 13]. PRF mixtures are representative of the spontaneous autoignition resistance in the conditions of RON (Research Octane Number-procedure D-2699 [14]) and MON (Motor Octane Number-procedure D-2699 [14]) experiments. However, RON and MON are not representative of all engine conditions of gasoline autoignition resistance [15]. Laminar flame speeds of PRF-mixtures are also different from gasoline laminar flame speeds [16, 17] and may induce errors during flame propagation simulations.

Toluene is a standard representative of aromatic compounds in commercial fuels, promising better predictions of autoignition delays in various conditions, since aromatics have a large octane number sensitivity (differences between RON and MON above 10) and different laminar flame speeds compared to alkanes. A three component blend including the binary Primary Reference Fuel (PRF) n-heptane/iso-octane and toluene has been chosen as the fuel surrogate and the interactions between HC and $NO_x$ were modeled.

Our modelling approach should be valid for the combustion in EGR equipped IC-engines, but should also describe post-oxidation phenomena in the exhaust line. Therefore a detailed kinetic mechanism was developed providing information about the fuel oxidation and the impact of $NO_x$. The modelling of the $NO_x$ interactions with HC is discussed in detail.

**RECENT WORK IN THE FIELD**

Published work in the field has been recently reviewed by Battin-Leclerc [18]. Amongst the different studies on the topic, we can mention here those of Minetti et al. [19, 20] Vanhove et al. [21], Callahan et al. [22] and Tanaka et al. [3], who have performed experiments on the autoignition behavior of different fuels in a Rapid Compression Machine (RCM) over a variety of thermodynamic conditions. Further, Ciezki and Adomeit [23], Fieweger et al. [24] and Gauthier et al. [25] have performed studies in Shock Tubes (ST) on the ignition of n-heptane, iso-octane [23, 24] and toluene [25] mixtures over a wide range of temperatures and pressures. Their experiments also covered a broad range of equivalence ratios and dilution gas proportions. Additionally, Andrae et al. [26] performed experiments in an HCCI-engine for PRF and n-heptane/toluene mixtures in various proportions. Detailed mechanisms [22], [17-29] , as well as reduced mechanisms have been developed [30, 31] to reproduce the oxidation of binary PRF fuel mixtures composed of n-heptane and iso-octane in variable proportions. Less detailed mechanisms have also been proposed for the oxidation of toluene/ PRF fuel mixtures, e.g. [26, 32].

There are numerous papers studying the effect of $NO_x$ on the oxidation of $C_1$-$C_2$ hydrocarbons [5], [23, 34] and of $C_3$-$C_5$ hydrocarbons [5], [7], [35, 38]. Concerning $C_3$-$C_5$ hydrocarbons, Dagaut et al. [37, 38] have investigated the NO reduction by propene and propane using a Jet-Stirred Reactor (JSR). This data was modelled by Frassoldati et al. [5]. Flow reactor experiments and detailed chemical kinetic calculations were performed by Hori et al. [35] to examine the promotion effect by NO on the oxidation of propane. Hori et al. [36] performed a similar study for n-butane and n-pentane. The investigated temperature range was 600-1200 K including the NTC temperature range. Glaude et al. [7] improved the modelling of the experimental data performed by Hori et al. [36].

Only a few studies are related to the impact of $NO_x$ on the oxidation of hydrocarbons containing more than 5 carbon atoms. Moréac [39] has studied the influence of the addition of NO on the oxidation of pure n-heptane, iso-octane and toluene using a JSR operated at 1 and 10 atm over a temperature range of 550 to 1180 K. A model was proposed to reproduce the results obtained for n-heptane [40]. Dubreuil et al. [4] provided experimental data for the oxidation of n-heptane/toluene and n-heptane/iso-octane mixtures with various amounts of added NO in a JSR for a pressure of 10 atm and in an HCCI-engine. They modelled the interactions between n-heptane/toluene mixtures and NO. The objective of this study is to extend the previous studies in proposing a detailed mechanism for the interaction of NO with ternary mixtures of n-heptane, iso-octane and toluene. Experimental data about the influence of the addition of NO on cool flame and full ignition delays in an HCCI engine were also published by Risberg et al. [41].

**DESCRIPTION OF THE PROPOSED MECHANISM**

The mechanism that we propose is based on the PRF autoignition model from Buda et al. [29] coupled with the model for the oxidation of toluene provided by Bounaceur et al. [42]. Additionally the reactions of $NO_x$ with PRF and toluene compounds were written. Pressure-dependent rate constants are defined by the formalism proposed by Troe [43]. Thermochemical data for molecules and radicals were calculated by the THERGAS software [44], which is based on additivity methods proposed by Benson [45]. It yields 14 polynomial coefficients according to the CHEMKIN formalism [46]. In the case of nitrogen containing compounds, thermochemical data proposed by Marinov [47] and Burcat [48] has been used. The new mechanism contains 3000 reactions and 536 species and will be available on request.

*Mechanism for the oxidation of n-heptane and iso-octane*

A mechanism for the oxidation of a n-heptane/iso-octane mixture has been generated by the software EXGAS-ALKANES according to the principles described by Buda et al. [29]. This mechanism is composed of three parts:

➢ A $C_0$-$C_2$ reaction base [49] involving species with up to two carbon atoms and including kinetic data mainly taken from the evaluations of Baulch et al. [50] and Tsang and Hampson [51]. To obtain a good agreement with the experimental results obtained at high pressures (above 40 atm), the rate constant of the decomposition of $H_2O_2$ defined in the $C_0$-$C_2$ reaction base had to be multiplied by a factor 4, compared to the rate constant defined for lower pressures. The need for this adjustment is probably related to a still unknown problem in the mechanism at high pressure.

➢ A comprehensive primary mechanism. It only considers initial organic compounds and oxygen as reactants. It includes all the usual low and intermediate temperature reactions of alkanes, which are:

- Unimolecular initiations involving the breaking of a C-C bond.
- Bimolecular initiations with oxygen to produce alkyl (R•) and hydroperoxy (•OOH) radicals.
- Additions of alkyl and hydroperoxyalkyl (•QOOH) radicals to an oxygen molecule.
- Isomerisations of alkylperoxy and hydroperoxyperoxy radicals (ROO• and •OOQOOH) involving a cyclic transition state, we consider a direct isomerization-decomposition to give ketohydroperoxides and hydroxyl radicals [52].
- Decompositions of radicals by β-scission involving the breaking of C-C or C-O bonds for all types of radicals (for low temperature modelling, the breaking of C-H bonds is not considered).
- Decompositions of hydroperoxyalkyl radicals to form cyclic ethers and •OH radicals.
- Oxidations of alkyl radicals with $O_2$ to form alkenes and •OOH radicals.
- Metatheses between radicals and the initial reactants (H-abstractions).
- Recombinations of radicals.
- Disproportionations of peroxyalkyl radicals with •OOH to produce hydroperoxides and $O_2$.

The kinetic rate constants used are those described by Buda et al. [29].

➢ A lumped secondary mechanism [53]. The molecules produced in the primary mechanism, with the same molecular formula and the same functional groups are lumped into one unique species without distinction between different isomers. The secondary mechanism includes global reactions producing, in the smallest number of steps, molecules or radicals whose reactions are included in the $C_0$-$C_2$ base.

Previous work on PRF fuels shows that co-oxidation reactions between n-heptane and iso-octane are negligible and that the coupling between their oxidation kinetics is based mainly on small radicals interactions [54] defined in the $C_0$-$C_2$ base.

*Mechanism for the oxidation of toluene*

The model for the oxidation of toluene includes the following sub-mechanisms [42]:

- A primary mechanism including reactions of toluene containing 193 reactions and including the reactions of toluene and of benzyl, tolyl, peroxybenzyl (methylphenyl), alkoxybenzyl and cresoxy free radicals.

- A secondary mechanism involving the reactions of benzaldehyde, benzyl hydroperoxyde, cresol, benzylalcohol, ethylbenzene, styrene and bibenzyl [42].

- A mechanism for the oxidation of benzene [55]. It includes the reactions of benzene and of cyclohexadienyl, phenyl, phenylperoxy, phenoxy, hydroxyphenoxy, cyclopentadienyl, cyclopentadienoxy and hydroxycyclopentadienyl free radicals, as well as the reactions of ortho-benzoquinone, phenol, cyclopentadiene, cyclopentadienone and vinylketene.

- A mechanism for the oxidation of unsaturated $C_0$-$C_4$ species. It contains reactions involving •$C_3H_2$, •$C_3H_3$, $C_3H_4$ (allene and propyne), •$C_3H_5$ (three isomers), $C_3H_6$, $C_4H_2$, •$C_4H_3$ (2 isomers), $C_4H_4$, •$C_4H_5$ (5 isomers), $C_4H_6$ (1,3-butadiene, 1,2-butadiene, methyl-cyclopropene, 1-butyne and 2-butyne).

In case of redundant reactions defined in both, the secondary mechanism generated for the alkanes oxidation and in the $C_0$-$C_4$ reaction base for the toluene oxidation (e.g. reaction of propene or of allyl radicals), we kept the reactions written in the n-heptane/iso-octane mechanism and skipped those defined in the toluene mechanism. The mechanism for the oxidation of toluene has been validated in previous work using experimental results obtained in JSR [42], plug flow reactors [56], and Shock Tubes (ST) [42].

*Cross-term reactions between alkanes and toluene*

In our model the following co-oxidation reactions between alkanes and toluene are defined:

- Metathesis of benzyl radicals with n-heptane and iso-octane leading to toluene and alkyl radicals (respectively heptyl and octyl) with an A factor of $1.6 \times T^{3.3}$ cm$^3$mol$^{-1}$s$^{-1}$ and an activation energy of 19.8, 18.2 and 17.2 kcal/mol for the abstraction of a primary, secondary and tertiary H-atom, respectively [57].
- Terminations between benzyl and alkyl radicals producing alkylbenzenes with a rate constant of $1.0 \times 10^{13}$ cm$^3$mol$^{-1}$s$^{-1}$ according to collision theory.
- Metathesis with toluene of secondary allylic radicals (iso-butenyl, iso-octenyl, heptenyl) with a rate constant of $1.6 \times 10^{12}\exp(-7600/T)$ cm$^3$mol$^{-1}$s$^{-1}$ as for allyl radicals [42] and of methyl peroxy (CH$_3$OO•) radicals with a rate constant of $4.0 \times 10^{13}\exp(-6040/T)$ cm$^3$mol$^{-1}$s$^{-1}$ [58].

*Reactions of NO$_x$ compounds*

The mechanism for the NO$_x$ species was derived from the modelling work of Glaude et al. [7] concerning the effect of the addition of NO on the oxidation of n-butane and n-pentane. This mechanism was based on the model of Hori et al. [35] on the conversion of NO to NO$_2$ promoted by methane, ethane, ethylene, propane, and propene. The kinetic model proposed by Glaude is primarily based on the GRI-MECH 2.11 [59] and the research performed by Dean and Bozzelli [60] and Atkinson et al. [61]. In this study we used the more recent mechanism GRI-MECH 3.0 [59]. Table 1 shows the reactions of NO$_x$ compounds which have been changed or added compared to the mechanisms of Glaude et al. [7] and GRI-MECH 3.0 [59] and the reactions written for NO$_x$ interacting with alkanes and aromatic compounds.

## TABLE I

Rate parameters of reactions (1, 2, 4-6, 8-12) were chosen from recent studies on reaction kinetics between C$_1$-C$_2$ species with nitrogen containing compounds [62]-[70] and are different from those used by Glaude et al. [7] for the same reactions. Rate parameters of reactions (3) and (7) have been adjusted in order to obtain satisfactory simulation results. The reactions of NO$_x$ with formaldehyde (HCHO) (reactions 5 and 6) and with nitrous acid (HONO) (reaction 12) and six reactions involving HNO radicals (reactions 7-11) were added with rate constants mainly taken from the literature.

Coupling reactions between species involved in the alkane oxidation model and $NO_x$ were written. These reactions were partly derived from the mechanism published by Glaude et al. [7] for the oxidation of n-butane and n-pentane in presence of NO. The reaction types are:

- H-abstractions from alkanes by $NO_2$ (reaction 14) with rate parameters proposed by Chan et al. [71].

- the reaction of alkyl radicals (R•) with $NO_2$ giving either $RNO_2$ (reaction 15) or alkoxy radicals (RO•) radicals (reaction 16) as it is written in the case of methyl radicals (•CH3) radicals (reactions 2 and 3).

The rate constant of reaction 16 was taken similar to that proposed by Glarborg et al. [72] for the corresponding reaction of •CH$_3$ radicals with $NO_2$.

- Reactions of ROO• radicals with NO forming $NO_2$ and RO• (reaction 17).

- Reactions of HOOQOO• radicals with NO which are decomposed via a global reaction producing $NO_2$, a hydroxyl radical (•OH), two HCHO molecules and the corresponding olefin (reaction 18).

The rate constants of reactions 17 and 18 were considered equal to the values proposed by Atkinson et al. [61] for the corresponding reaction of methylperoxy radicals ($CH_3OO$•) with NO.

- Decomposition of RO• radicals by beta-scission (reaction 19) with the rate constant proposed by Curran et al. [73].

- H-abstractions from aldehydes by $NO_2$ (reaction 20) with rate parameters estimated as in the case of formaldehyde (reaction 6).

- Reactions of resonance stabilized allylic radicals with $NO_2$ yielding NO, acrolein and an alkyl radical (R•) (reaction 21) with the rate constant proposed by Glaude et al. [7].

Sensitivity analysis allowed us to neglect other reactions of alkoxy radicals, which were considered by Glaude et al. [7]. In contrast, the reactions of alkane molecules (reaction 14) and alkyl radicals (reactions 15-16) with $NO_2$ were added, while they were not considered by Glaude et al. [7].

For the coupling of nitrous species with aromatic compounds the following reactions were written with rate constants which had to be estimated in most cases because no corresponding kinetic data was available:

- H-abstractions from benzene by $NO_2$ and NO (reactions 22-23) with rate constant proposed by Chan et al. [71].

- Reactions of phenyl radicals with HNO (reactions 24-25), $NO_2$ (reaction 26) and NO (reaction 27). Rate constants were taken from recent papers in the literature [68], [74], [75].

- Reactions of phenylperoxy (reaction 28) and benzylperoxy (reaction 35) radicals which were derived from analogous reactions of alkylperoxy radicals.

- H-abstractions from toluene by $NO_2$ and NO (reactions 29-32).

- Reactions of resonance stabilized benzyl radicals with $NO_2$ (reactions 33-34).

- Reactions of benzylalkoxy (reactions 36-38) and benzaldehyde, (reactions 39 and 40) with rate constants derived from the corresponding reactions of methoxy radicals and formaldehyde, respectively.

- Reactions $HOC_6H_4CH_2$ with nitrogen containing species (reactions 41-43) with rate constants derived from the corresponding reactions of benzyl radicals.

- Reactions of $C_6H_5CH_2NO_2$ with small radicals (reactions 44-48) with rate constants derived from the corresponding reactions of $CH_3NO_2$.

- Reactions of $C_6H_5NO_2$ radicals with rate constants taken from Xu et al. [74] (reaction 49) and Park et al. [76] (reaction 50).

# COMPARISON BETWEEN SIMULATED RESULTS AND PREVIOUSLY PUBLISHED EXPERIMENTAL DATA

Validations were performed over a wide range of thermodynamic conditions for different surrogate fuel compositions in various experimental setups. All simulations were run with the commercial software code CHEMKIN IV [46].

## Oxidation of PRF and toluene containing fuels in absence of $NO_x$

Our PRF/toluene/$NO_x$ oxidation model is first tested without $NO_x$ interactions. We present here validations performed against autoignition delay data obtained in RCM, ST, and HCCI engines. Further we provide validations for species profiles measured for slow oxidations in highly diluted JSR. The summaries of the experimental conditions used for the different facilities are presented in Tables II, III and IV, respectively.

### TABLES II, III, IV

*Rapid Compression Machines (RCM) and Shock Tubes (ST)*

Figures 1 and 2 compare experimental and simulated results for RCM experiments performed by Vanhove et al. [21]. The authors measured both cool flame ($t_{cf}$) and main ignition ($t_{ig}$) delay times for pure n-heptane, pure iso-octane, n-heptane/toluene and iso-octane/toluene blends in a RCM as a function of temperature. They obtained temperature variations in a range from 600 to 900 K at the end of compression by varying the composition of the inert gas. The temperature obtained at the end of the compression was calculated using the adiabatic core gas model. The authors deduced ignition and cool flame delay times from the pressure and light-emission traces.

Their experiment for the oxidation of pure n-heptane (Fig. 1a) and iso-octane (Fig. 1b) reveals that at temperatures around 800 K there is a zone with increasing ignition delays associated to increasing temperatures, which is characteristic of the NTC-regime. Compared to the delays measured for the oxidation of pure n-heptane and iso-octane, the addition of toluene to n-heptane (Fig. 2a) and to iso-octane (Fig. 2b) increases the ignition delays and reduces the NTC-effect at intermediate temperatures, .

**FIGURES 1 AND 2**

Our simulations show the ignition delays of both cool and hot flames produced by the mechanism are qualitatively correct. However, in the case of pure iso-octane (Fig.1b), cool flame delays and main ignition delays above 800 K are overestimated. The retarding impact of toluene on the n-heptane and iso-octane oxidation is retrieved (Fig. 2), showing stronger toluene interactions in the case of iso-octane. For the oxidation of toluene/n-heptane mixtures (Fig. 2a) the mechanism overestimates ignition delays below 700K and underestimates them at above 800K (Fig. 2a). Nevertheless, the appearance of cool flames and main flame ignition delays are qualitatively well captured. Compared to experiments, simulations predict generally shorter ignitions delays for the oxidation of iso-octane/toluene mixtures (Fig. 2b). This is believed to be caused by an insufficient coupling of iso-octane/toluene for capturing the retarding impact of toluene on the iso-octane oxidation.

Callahan et al. [22] measured main flame ignition delays for n-heptane/iso-octane mixtures at pressures between 11 and 17 atm by varying the RON number, characterised by the iso-octane/n-heptane ratio. Figure 3 displays measured and simulated main flame ignition delays for a RON of 100, 95 and 90. Experiments and simulations show decreasing ignition delays for decreasing RON. The simulations qualitatively retrieve ignition delays over the analyzed temperature range, including the NTC effect, but generally ignition delays are slightly overestimated. This is especially the case for PRF 90 where the accelerating impact of n-heptane on the iso-octane ignition is reproduced not well reproduced. This is similar to the results obtained for the oxidation of pure iso-octane presented in Figure 1b.

**FIGURE 3**

Ciezki et al. [23] and Fieweger et al. [24] measured ignition delay times with ST experiments at around 13 atm and varied the equivalence ratio for the oxidation of pure n-heptane. They interpreted ignition delays by visualizing CH emissions and interpreting pressure signals with an accuracy of around ±20μs. Figure 4 shows the comparison between measured data and simulated results. Experiments and simulations both show decreasing ignition delay times for increasing equivalence ratios. Generally, ignition delays for n-heptane oxidations are captured correctly, even if at low temperatures the model tends to overpredict ignition delays.

**FIGURE 4**

Fieweger et al. [24] performed ST experiments for the stoichiometric oxidation of PRF-mixtures at pressures around 40 atm. They varied the RON number of the investigated PRF mixtures from 0 to 100. Experimental data and simulation results are displayed in Figure 5. Experiments show a significant impact of the RON number on ignition delays for a temperature range between 600 and 1200 K, which is reproduced well by the model.

**FIGURE 5**

Tanaka et al. [3] have studied the autoignition behavior of different fuels in a RCM. Their study focuses on the sensitivity of ignition delay times to small variations on the iso-octane/n-heptane ratio with a fixed amount of toluene. They obtained mixture purities greater than 99 %, recorded the pressure histories and deduced temperature evolutions by using isentropic relations. Fractional errors in the recorded peak pressures and ignition delay times are declared less than ±3 %. Recorded results are compared to simulations for n-heptane/iso-octane/toluene mixtures and are shown in Figure 6. Experimental results of all investigated fuel blends reveal the appearance of a cool flame, what is correctly reproduced by simulations. The stepwise increase of the iso-octane concentration causes a non-linear increase of the main ignition delay which is also correctly captured by the model.

**FIGURE 6**

Gauthier et al. [22] measured ignition delay times of ternary n-heptane/iso-octane/toluene mixtures in a ST. They performed experiments at medium (15-20 atm) and high pressures (45-60 atm), varying temperatures from 850 to 1100 K. In the considered temperature range the obtained discrepancies of recorded delay times are indicated by the authors as being less than 10μs. Their study examines the sensitivity of ignition delay times by varying the toluene/iso-octane ratio with a constant n-heptane concentration. The comparison between experimental data and simulation results is shown in Figure 7. Experiments show that an increase of pressure reduces the ignition delays. At high pressures, ignition delays become quasi independent of the toluene/iso-octane ratio (compare Fig 7a and Fig 7b), which is not the case at lower pressures. In the pressure range from 15 to 20 atm and for an increased toluene/iso-octane ratio retarded ignition delays are observed. The described characteristics are well reproduced by the model. However an overprediction of ignition delays is generally observed in the investigated temperature and pressure range.

**FIGURE 7**

*HCCI engines*

Validations in HCCI engines are of importance because they allow the model to be tested on real problems for which it was designed. Andrae et al. [58] have performed experiments in an HCCI-engine with PRF and n-heptane/toluene mixtures. They performed experiments for 2 different experimental set-ups varying equivalence ratios, initial engine temperatures and initial engine pressures. The engine was operated at 900 rpm and the pressure as a function of Crank Angle Degree (CAD) After Top Dead Center (ATDC) in the HCCI-engine was measured. Figure 8a shows experimental and simulated results obtained for PRF mixtures characterized by two different RON. Two different engine runs (Run_1 and Run_2 in Table III) were tested varying admission temperature and equivalence ratio. Experimental and simulated data are shown for Fuel_Andr_1 (6 % n-heptane, 94 % iso-octane) for the engine operation conditions Run_1 and Run_2 and compared to the engine combustion of Fuel_Andr_2 for engine operation condition Run_2. Figure 8b presents the analogous comparison between experimental and simulated data for the two different toluene/n-heptane mixtures (Fuel_Andr_3, Fuel_Andr_4).

Experimental results show that lowering the iso-octane/n-heptane ratio (Figure 8a) and decreasing the toluene/n-heptane-ratio (Figure 8b) results in an increased reactivity and an advanced fuel ignition. This characteristic is well captured by the model. In addition the impact of pressure and temperature variations leading to an advanced fuel ignition for the engine set-up Run_2 is correctly predicted. The overprediction of maximum pressures is due to the simplifying assumption of an adiabatic combustion and a perfectly stirred fuel/air mixture.

**FIGURE 8**

*JSR*

Validations in JSR are of importance because they allow the model to be tested for the consumption and formation of different species. Moréac et al. [39] studied the oxidation of n-heptane, iso-octane and toluene at different pressures for temperatures from 500 to 1100 K. They recorded reactant mole fractions (including NO and $NO_2$) by using chromatographic techniques with uncertainties of ±10 % except for toluene and benzene, where the uncertainties reached up to ±15 %.

Figures 9 and 10 compare experimental and simulated results for the oxidation of n-heptane at 1 atm and 10 atm, respectively. At atmospheric pressure no n-heptane conversion (Fig. 9a) and no production of CO (Fig. 9b) is observed below 900 K. Simulations satisfactorily reproduce the non-reactivity of n-heptane at low temperatures. At 10 atm pure n-heptane starts reacting at temperatures around 520 K (Fig. 10a) associated to a CO production (Fig. 10b), which is modelled correctly. However, simulations underestimate the strong NTC behavior observed experimentally and do not reproduce the very slight inhibiting effect between 700 and 800 K shown in experiments.

**FIGURE 9 AND FIGURE 10**

Figure 11 compares experimental data of Moréac et al. [39] to simulated results for the stoichiometric oxidation of iso-octane at atmospheric pressure. As for n-heptane, no reactant conversion (Fig. 11a) and no CO production (Fig. 11b) is observed in experiments and simulations.

**FIGURE 11**

Experimental results of Moréac et al. [39] and simulations for the oxidation of toluene at a pressure of 10 atm are shown in Figure 12. Experiments reveal that toluene starts reacting at temperatures around 900 K, which causes a strong diminution of the toluene concentration (Fig. 12a) and a strong increase of that of CO (Fig. 12b). These characteristics are correctly captured by the model.

**FIGURE 12**

Dubreuil et al. [4] performed experiments in a JSR for the oxidation of PRF mixtures and for n-heptane/toluene mixtures at a pressure of 10 atm. The authors indicate a fuel purity of greater than 99.9 %, while they obtained variations in the carbon balance of measured species of around ±5 %. The recorded iso-octane concentrations as a function of temperature and addition of NO show similar characteristics as those obtained by Moréac et al. [39] for the oxidation of n-heptane at identical pressure. The comparison of measured and calculated iso-octane concentrations presented in Figure 13 shows good agreement with experimental observations.

**FIGURE 13**

The experimental and computed data for the oxidation of n-heptane/toluene mixtures are shown in Figure 14. The temperature at which the n-heptane/toluene mixture react is around 580 K which corresponds to the value measured by Moréac et al. [39] for the oxidation of pure n-heptane. Under similar conditions, pure toluene only starts to react at around 900 K [39] indicating the promoting effect caused by the presence of n-heptane on the toluene oxidation. A satisfactory agreement between simulation and experimental results is obtained.

FIGURE 14

**Oxidation of PRF and toluene fuels in presence of $NO_x$**

We present here the validations of our mechanism against experiments investigating the impact of $NO_x$ on the oxidation of HC. Experimental data of ignition delays were obtained in HCCI engines and species profiles in a highly diluted JSR. A summary of the considered experimental conditions is shown in Tables III and IV, respectively. It is worth noting that no experimental data obtained in RCM or ST is available, which investigates the impact of the addition of NO on autoignition.

*HCCI engines*

Dubreuil et al. [4] performed experiments for the oxidation of pure n-heptane, PRF and n-heptane/toluene mixtures with the addition of various concentrations of NO for a test-engine running at 1500 rpm. They observed the evolution of cool- flame and main flame ignition delays as a function of the added amount of NO. Ignition delays were obtained by calculating the heat release rate from the recorded pressure history as a function of the Crank Angle Degree (CAD). It should be mentioned that at the investigated engine speed of 1500 rpm the engine crank turns 1 angle degree in around 0.1 ms.

FIGURE 15

Figure 15 shows the effect of NO addition on cool flame ignition delays (Fig. 15a) and main flame ignition delays (Fig. 15b) for the three fuels referenced in Table III (Fuel_Dub_1, Fuel_Dub_2, Fuel_Dub_3). Experiments show that cool flame ignition delays decrease with increasing NO concentrations up to NO concentrations of around 100 ppm. For higher concentrations of added NO, the ignition delay increases up to the maximum concentration of 500 ppm without reaching the initial ignition delay obtained in absence of NO. In contrast, the main flame ignition delay decreases strongly for a small addition of NO (up to 50 ppm) while it remains constant for higher NO concentrations. The model shows only slight sensitivity of both, cool flame and ignition delays to the addition of NO (Fig 15a). However, the retarding impact on the ignition delays by blending pure n-heptane (Fuel_Dub_1) with iso-octane (Fuel_Dub_2) and toluene (Fuel_Dub_3) is correctly reproduced (Fig 15b). Cool flame and main flame ignition delays are overestimated by less than 10 %.

Risberg et al. [41] performed experiments on the oxidation on pure PRF and n-heptane/toluene mixtures with the addition of various concentrations of NO for a test-engine running at 900 rpm. They traced the heat release evolution as a function of CAD during cool flame and main flame ignition. The heat release rate was obtained from the measured pressure evolution as a function of the added amount of NO.

**FIGURE 16**

Figure 16 shows the effect of NO addition on cool flame ignition delays (Fig. 16a) and main flame ignition delays (Fig. 16b) for the two fuels referenced in Table III (Fuel_Ris_1, Fuel_Ris_2). In both cases, the ignition delay time has been deduced from the heat release profile by taking the CAD value at half of the reached maximum value of the heat release. In the case of the PRF-mixture, as well as in the case of the n-heptane/toluene mixture, the experiments show a decrease of the cool flame ignition delays when the NO concentrations are increased up to 75 ppm. For this NO concentration, the cool flame ignition delays reach a minimum and are retarded for higher NO contents. This is not the case for the main flame ignition delays which decrease over the complete range of tested NO concentrations.

To model these experimental results, we adjusted the initial temperature such that ignition delays in absence of NO were correctly reproduced. We simulated then the impact of NO addition by keeping constant initial thermodynamic conditions. These procedure corresponds to a validation of the model sensitivity against NO addition.

Our model captures well the effect of the NO addition on the cool flame and main flame ignition delays for both fuels. Nevertheless, the model predicts a higher sensitivity to NO for PRF-mixtures than for n-heptane-/toluene mixtures which is in contrast with experimental results.

*JSR*

Moréac et al. [39] also studied the impact of NO on the oxidation of n-heptane, iso-octane and toluene at different pressures and at temperatures from 500 to 1100 K. Figures 9 and 11 show that, at atmospheric pressure, the experimentally observed promoting impact of NO on both the conversion of reacting alkane (Fig. 9a and 11a) and the CO concentration profiles (Fig. 9b and 11b) are captured well by the model. Experimental results presented in Figure 10 for the oxidation of n-heptane at a pressure of 10 atm reveal a complex impact of NO addition. The addition of NO in small concentrations (50 ppm) shows a slight retarding effect on the minimum temperature at which n-heptane starts to react. In contrast, the addition of larger amounts of NO (500 ppm) causes a strong inhibition and the minimum temperature above which n-heptane reaction is detected shifts up to 650 K. One also observes that the NTC-effect between 650 and 750 K is reduced by small amounts of added NO (50 ppm) and completely disappears, when NO is present in larger concentrations (500 ppm). Above 750 K and at increased pressure (10 atm), the addition of NO generally promotes the oxidation of n-heptane stronger than at atmospheric pressure. Neither the inhibition of the n-heptane oxidation at low temperatures, nor the acceleration by NO at higher temperatures is linear compared to the amount of added NO.

The comparison of the simulation results with the experimental data shows that the model predicts the inhibiting impact of NO at low temperature, as well as its promoting effect at higher temperature. In addition the reduction of the NTC effect for small concentrations of NO (50 ppm) and the sudden decrease of the n-heptane concentrations for the addition of 500 ppm of NO at 650 K are captured well by the model.

The experimental results for the oxidation of toluene displayed in Figure 12 show that the addition of small amounts of NO (50 ppm) leads to a shift of the toluene and CO concentration profiles towards lower temperatures by 30 to 50 K. An addition of 500 ppm of NO results in further shifts of the toluene and CO concentration profiles of 40 to 60 K and the acceleration of the toluene oxidation by NO is non-linear with respect to the amount of NO added. The kinetic model captures well the shift in toluene and CO concentration profiles, but the reactivity of toluene in presence of NO is underestimated.

Dubreuil et al. [4] also performed as well experiments in a JSR for different fuel mixtures and several contents of NO. Variations of recorded NO and $NO_2$ concentrations are indicated between ±5-20ppmv. As shown in Figure 13, the impact of NO on the PRF oxidation observed experimentally is similar to that obtained by Moréac et al. [39] for the oxidation of pure n-heptane at 10 atm. The addition of NO inhibits the oxidation of PRF mixtures below 700 K, while it promotes it at higher temperatures. The NTC-effect around 750 K is reduced by the addition of small amounts of NO (50 ppm) and disappears completely when NO is added in higher concentration (500 ppm). The comparison of experimental results to the simulation shows that the described effects of NO addition are well captured by the model. However, for high concentrations of NO (200 ppm) the model overestimates the minimum temperature above which the reactivity of iso-octane is detected.

Experimental results presented in Figure 14 for the oxidation of n-heptane/toluene show, that the addition of NO causes an inhibition of the n-heptane/toluene oxidation at temperatures below 700 K, while at higher temperatures a promoting effect is observed. These effects either are reproduced well by the model.

# SENSITIVITY AND FLUX ANALYSYS FOR NITROUS COMPOUNDS CONTAINING REACTIONS

Sensitivity and reaction rate analyses were performed with the previously described mechanism for JSR-simulations of neat n-heptane and toluene oxidation for various temperatures (665 K, 750 K, 950 K), different pressures (1 atm, 10 atm) and two amounts of added NO (50 ppm, 500 ppm). The simulation conditions correspond to the experimental conditions chosen by Moréac et al. [39] for the previously shown experiments on the oxidation of pure HC in a JSR (Table IV).

The relative mol reaction rates indicated in Figure 16 and Figure 20 for reaction "x" represent the mol flows via reaction "x" normalized by the rate of production of heptyl by benzyl radicals, respectively. Sensitivity coefficients ($\sigma$) were obtained by the following formula:

$$\sigma_x = \frac{M_{reactants}(k_x \times 10) - M_{reactants}(k_x/10)}{9.9 \times M_{reactants}(k_x)},$$

where $M_{reactant}(k_x \times 10)$ is the reactant mole fraction obtained for a simulation run with the rate constant of reaction x multiplied by a factor 10, $M_{reactant}(k_x/10)$ the reactant mole fraction obtained for a simulation run with the rate constant of reaction x divided by a factor 10 and $M_{reactant}(k_x)$ the reactant mole fraction simulated by the initial mechanism. A positive sensitivity coefficient $\sigma_x$ indicates an inhibitive effect and a negative coefficient shows an accelerating impact of reaction (x) on the global reactivity.

*Oxidation of n-heptane*

Figures 16 to 18 show the reaction rate and sensitivity analyses performed for the oxidation of n-heptane.

**FIGURE 16, FIGURE 18, FIGURE 19**

Figure 16a shows the classical scheme for the oxidation of an alkane at low temperatures (665 K) in absence of $NO_x$. The main reaction channel is the formation of alkyl (R•) radicals from the initial reactant followed by an addition of oxygen molecules and an isomerisation of the obtained peroxy radicals (ROO•) to give hydroperoxyalkyl (•QOOH) radicals. These radicals can decompose into stable species, such as cyclic ethers or ketones, involving the expulsion of •OH radicals. They can also react by the addition with another oxygen molecule producing hydroperoxyalkylperoxy (•OOQOOH) radicals. The isomerisation and decomposition of •OOQOOH radicals lead to the formation of hydroperoxide molecules. The composition of these hydroperoxide molecules involve a multiplication of the radical production, which in a chain reaction induces an exponential acceleration of the reaction rate. At temperatures around 750 to 800 K, the reversibility of the addition of alkyl (R•) radicals to oxygen molecules becomes more important. The oxidation of these radicals leading to the formation of alkenes and the very unreactive •OOH radicals is then favoured. This reduces the overall reactivity and is the main reason for the appearance of the NTC regime [29].

The presence of $NO_x$ considerably changes the reaction scheme of the oxidation of alkanes. This is illustrated in Figures 16b and 16c. The reaction of peroxy radicals with NO gives alkoxy (RO•) radicals and $NO_2$ (reaction 17 in Table I). The resulting RO• radicals are decomposed to aldehydes and smaller alkyl radicals (reaction 19). This reaction channel competes with the second addition of oxygen molecules and thus reduces the rate of formation of hydroperoxide species disadvantaging the branching steps which are induced by the decomposition of •OOQOOH radicals. With an increased amount of added NO a considerable rise of the RO• production is observed. For an addition of 500 ppm of NO at 10 atm and 665 K, 96 % of peroxy radicals are converted to alkoxy radicals RO• (Fig. 16c). This explains the inhibiting impact of $NO_x$ on HC oxidation between 650 and 750 K. The sensitivity analyses displayed in figures 17 and 18 reveals that reaction (17) has an inhibiting impact on the n-heptane oxidation under any investigated conditions. This inhibition is particularly important at 650 K and a pressure of 10 atm. It can be noted that the reactions of hydroperoxyalkylperoxy radicals (HOOQOO•) and NO are always of negligible importance and could have been omitted.

At low temperatures, a small proportion of alkyl radicals may react with $NO_2$ yielding $RNO_2$ molecules (reaction 15) or alkoxy radicals and NO (reaction 16). The production of $RNO_2$ molecules is mainly important at atmospheric pressure and temperatures below 750 K, where it strongly inhibits the global reactivity thus compensating the accelerating impact of the reactions of $CH_3OO•$ (reaction 52) and $•CH_3$ (reaction 2) with $NO_x$ (Fig. 17a). At higher pressures (10 atm) the production of alkoxy radicals by reaction (16) has also a noticeable effect. This reaction competing with the addition to oxygen molecules has an inhibiting effect, which is particularly important at 10 atm and 750 K (Fig. 18b). The reaction between alkanes and $NO_2$, leading to nitrous acid (HONO) (reaction 14) has a slight impact at 1 atm and 665K.

The H-abstractions from aldehydes by $NO_2$ (reaction 20) and the reaction of resonance stabilized (Y•) radicals with $NO_2$ (reaction 21) producing acrolein, NO and alkyl radicals show only a limited effect on the overall reactivity. Above the NTC zone (900 K), reactions of alkyl radicals with oxygen molecules become less important and the influence of reactions (14-21) is almost negligible.

Sensitivity analyses further show that the reactions of $NO_x$ with $C_1$-$C_2$ species are of great importance whatever the temperatures range. Under almost any condition of HC oxidation, reactions (2), (51), (52) and (53) show large sensitivity coefficients:

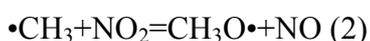
•$CH_3$+$NO_2$=$CH_3O$•+NO (2)

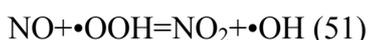
NO+•OOH=$NO_2$+•OH (51)

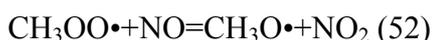
$CH_3OO$•+NO=$CH_3O$•+$NO_2$ (52)

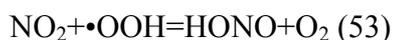
$NO_2$+•OOH=HONO+$O_2$ (53)

The promoting reactions (2) and (51) are particularly important at atmospheric pressure and at low temperatures, where the mechanism is very sensitive to the addition of NO. At low temperatures, the general fate of •OOH radicals is their recombination to form $H_2O_2$. Methyl radicals (•CH3) lead to the production of rather stable $CH_3OO•$ radicals via the addition of $O_2$. Reaction (51) transforms the unreactive •OOH into the very reactive •OH radicals, while reaction (2) produces reactive $CH_3O•$ radicals from •$CH_3$ radicals, thus increasing the overall reactivity. Reaction (53) is a termination step competing with reaction (51) showing generally an inhibiting effect. The impact of the reaction of $CH_3OO•$ radicals with NO (reaction 52) is more complex. At 10 bar it inhibits the global reactivity at 665K, but accelerates it at temperatures above 900 K (Fig. 18). The inhibiting impact of reaction (52) at low temperature might be caused by its competition with the disproportionation of $CH_3OO•$ with •OOH radicals yielding the branching agent, $CH_3OOH$, and oxygen molecule. At higher temperatures, this disproportionation becomes less important and thus reaction (52) has an accelerating impact. It should be noted that the contribution of the recombination of •$CH_3$ radicals and $NO_2$ via reaction (3) is negligible compared to the impact of larger alkyl radicals reacting via the analogous reaction type (15) (Fig. 17a).

Below 650 K, at a pressure of 10 atm and for a high concentration of NO (500 ppm), the alkane oxidation is governed by reaction (1).

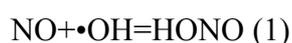

NO+•OH=HONO (1)

This explains the strong inhibition of n-heptane oxidation in presence of NO at temperatures below 650 K shown by Figure 9a. At these temperatures, reactive •OH radicals are consumed via reaction (1) producing nitrous acid (HONO) and representing thus an important sink for •OH radicals. At temperatures above 650 K, the dissociation of HONO exceeds its production and the sense of reaction (1) is reversed. The reverse of sense of reaction (1) causes a sudden acceleration of the overall reactivity and provokes the sudden decrease of the reactant concentration profile shown in Figure 9a.

One other important reaction is reaction (13) describing the reaction of ethyl ($C_2H_5\bullet$) radicals with $NO_2$. At low temperature, in absence of NO, the main reactions of ethyl radicals are with oxygen molecules to give the rather unreactive $C_2H_5OO\bullet$ radical. Reaction (13) provides an alternative channel toward the more reactive ethoxy ($C_2H_5O\bullet$) radical, explaining thus their promoting effect. Above 800 K, reaction (13) becomes inhibiting: at these temperatures, $C_2H_5\bullet$ radicals react mainly with $O_2$ to give ethylene and $\bullet OOH$ radicals, which again are transformed to $\bullet OH$ radicals via reaction (51). The competing reaction (13) thus has an inhibiting impact, as it produces stable ethoxy radicals instead of the easily converted $\bullet OOH$ radicals.

*Oxidation of toluene*

Figure 20 and Figure 21 show, respectively, the reaction rates and sensitivity analyses performed for the oxidation of toluene at 900 K in the case of the reaction of aromatic compounds. Figure 20 shows the sensitivity coefficients of governing reactions containing aromatic compounds and $NO_x$.

**FIGURE 20 AND FIGURE 21**

Bounaceur et al. [42] showed that at around 900 K, in absence of NO, toluene is mainly consumed to give resonance stabilized benzyl radicals. Thus, all relative mol fluxes indicated in Figure 20 are normalized to the rate of production of benzyl radicals. Benzyl radicals mainly react by combination with themselves to give bibenzyl or with $\bullet OOH$ radicals producing benzylhydroperoxide molecules and rapidly yielding alkoxybenzyl radicals which decompose to produce benzaldehyde and $H\bullet$ atoms or formaldehyde and phenyl radicals. A minor channel involves the reaction of benzyl radicals with oxygen molecules to give peroxybenzyl radicals, which isomerize and decompose to give benzaldehyde and $H\bullet$ atoms.

In the presence of nitrogen containing species, 5 % of toluene is consumed by H-abstractions with $NO_2$ producing HONO and benzyl radicals (reaction 30). Due to the rapid dissociation of HONO (reaction 1) giving $\bullet OH$ radicals and NO, this reaction has an important promoting effect in the studied temperature range, even if it leads to the formation of resonance stabilized benzyl radicals. Benzyl radicals can react with $NO_2$ to give directly alkoxybenzyl radicals and NO (reaction 33) and this reaction shows an important promoting effect. The combination of benzyl radicals with $NO_2$ (reaction 34) competes with reaction (33) and has an inhibiting effect. Peroxybenzyl radicals, which tend not to easily isomerize, can react with NO and produce alkoxybenzyl radicals (reaction 35) promoting the production of alkoxybenzyl. Nevertheless, in the presence of $NO_x$, the importance of reaction (33) is determinant: According to Figures 19b and 19c more than 90 % of benzyl radicals are consumed via this channel. In comparison, the effect of reaction (35) is rather limited, as less than 1 % of peroxybenzyl radicals are consumed via this reaction channel. One observes that the H-abstractions from benzaldehyde by $NO_2$ (reaction 39) has a promoting effect, while the reaction of peroxyphenyl with NO producing resonance stabilized phenoxy radicals and $NO_2$ (reaction 28) has an inhibiting impact.

**CONCLUSION**

The kinetic model for the oxidation of a PRF/toluene blend presented here has been successfully validated against different experimental applications over a wide range of thermochemical conditions. Ignition delays obtained in RCM, ST and HCCI experiments and concentration profiles measured in JSR experiments have been modelled.

A model describing the impact of $NO_x$ on HC oxidation has been developed and coupled with the PRF/toluene mechanism. This model was validated against HCCI and PSR experiments for neat fuel and various fuel blends. Validations show that the model accurately captures the complex impact of NO on HC oxidation. The impact at different pressures and varying temperatures for various concentrations of NO is well retrieved for all tested fuels. The sub-model containing nitrogen species was analysed by sensitivity and flux analyses and the important reaction channels have been identified permitting a deeper understanding of the impact of $NO_x$ on HC oxidation. The good results of model validations for different experimental setups over a wide range of thermochemical conditions should allow the use of the proposed mechanism for IC-engine applications, as well as for post-oxidation applications governing gas flows in the exhaust line.

**ACKNOWLEDGMENTS**

The authors wish to thank Dr. P.A. Glaude for providing the mechanism for the oxidation of n-pentane in presence of NO of Glaude et al. [7].

**TABLE I: MODIFIED REACTIONS OF $NO_x$ COMPOUNDS AND CROSS-TERM REACTIONS INVOLVED BY THE ADDITION OF NO TO THE OXIDATION OF ALKANES AND TOLUENE**

The rate constants are given ($k = A\, T^n \exp(-E_a/RT)$) in cc, mol, s, cal units. Reference numbers are given in brackets when they appear for the first time.

| Reactions | A | n | $E_a$ | References | No |
|---|---|---|---|---|---|
| **Reactions modified or added compared to Glaude et al. [7] or to GRI-MECH 3.0 [59]** | | | | | |
| NO+•OH=HONO (high pressure) | $1.1 \times 10^{14}$ | -0.3 | 0.0 | Atkinson04 [62] | (1.) |
| (low pressure) | $2.35 \times 10^{23}$ | -2.4 | 0.0 | | |
| Fall off parameter : $F_c$ =0.81 | | | | | |
| •$CH_3$+$NO_2$=$CH_3O$•+NO | $1.36 \times 10^{13}$ | 0.0 | 0.0 | Srinivasan05 [63] | (2.) |
| $CH_3NO_2$=•$CH_3$+$NO_2$ (high pressure) | $1.8 \times 10^{17}$ | 0.0 | 58500 | Estimated[a] | (3.) |
| (low pressure) | $1.3 \times 10^{18}$ | 0.0 | 42000 | | |
| $CH_3O$•+NO=HCHO+HNO | $7.6 \times 10^{13}$ | -0.76 | 0.0 | Atkinson05 [64] | (4.) |
| HCHO+NO=•CHO+HNO | $1.02 \times 10^{13}$ | 0.0 | 40670 | Tsang91 [65] | (5.) |
| HCHO +$NO_2$= •CHO+HONO | $8.35 \times 10^{-11}$ | 6.68 | 8300 | Xu03 [66] | (6.) |
| HNO+$O_2$=•OOH+NO | $8.0 \times 10^{10}$ | 0.0 | 9520 | Estimated[b] | (7.) |
| HNO+$NO_2$=$HNO_2$+NO | $3.0 \times 10^{11}$ | 0.0 | 1988 | Tsang91[65] | (8.) |
| HNO+NO=•OH+$N_2O$ | $8.5 \times 10^{12}$ | 0.0 | 29640 | Diau95 [67] | (9.) |
| HNO+•$CH_3$=$CH_4$+NO | $1.47 \times 10^{11}$ | 0.76 | 349.0 | Choi05 [68] | (10.) |
| HNO+$CH_3O$=$CH_3OH$+NO | $3.16 \times 10^{13}$ | 0.0 | 0.0 | He88 [69] | (11.) |
| HONO+NO=$NO_2$+HNO | $4.40 \times 10^{03}$ | 2.64 | 4038 | Mebel98 [70] | (12.) |
| $C_2H_5$• +$NO_2$=$C_2H_5O$•+NO | $1.36 \times 10^{13}$ | 0.0 | 0.0 | Estimated[c] | (13.) |
| **Oxidation of alkanes** | | | | | |
| RH+$NO_2$=R•+HONO | $\alpha \times 2.2 \times 10^{13}$ | 0.0 | 31100 | Chan01[d] [71] | (14.) |
| | $\beta \times 5.8 \times 10^{12}$ | 0.0 | 28100 | | |
| | $\gamma \times 9.3 \times 10^{13}$ | 0.0 | 25800 | | |
| $RNO_2$= R•+$NO_2$     (high pressure) | $1.8 \times 10^{17}$ | 0.0 | 58500 | Estimated[e] | (15.) |
| (low pressure) | $1.3 \times 10^{18}$ | 0.0 | 42000 | | |
| R•+$NO_2$ => RO•+NO | $4.0 \times 10^{13}$ | -0.2 | 0.0 | Estimated[f] | (16.) |
| ROO•+NO => RO•+$NO_2$ | $4.70 \times 10^{12}$ | 0.0 | -358.0 | Estimated[g] | (17.) |
| •OOQOOH+NO => •OH+2HCHO+olefin +$NO_2$ [h] | $4.70 \times 10^{12}$ | 0.0 | -358.0 | Estimated[h] | (18.) |
| RO• =>  aldehyde+ R'• | $2.0 \times 10^{13}$ | 0.0 | 15000 | Curran98 [72] | (19.) |
| aldehyde+$NO_2$=>R•+CO+HONO | $8.35 \times 10^{-11}$ | 6.68 | 8300 | Estimated[i] | (20.) |
| Y•+$NO_2$ => acrolein + R'•+NO[h] | $2.35 \times 10^{13}$ | 0.0 | 0.0 | Glaude05 [28] | (21.) |
| **Oxidation of benzene and toluene** | | | | | |
| **Reactions of benzene and phenyl radicals** | | | | | |
| $C_6H_6$+$NO_2$=$C_6H_5$•+HONO | $7.41 \times 10^{13}$ | 0.0 | 38200 | Chan01 [71] | (22.) |
| $C_6H_6$+$NO_2$=$C_6H_5$•+$HNO_2$ | $2.5 \times 10^{14}$ | 0.0 | 42200 | Chan01 [71] | (23.) |
| $C_6H_5$•+HNO=$C_6H_6$+NO | $3.78 \times 10^{5}$ | 2.28 | 456 | Choi05 [68] | (24.) |
| $C_6H_5$•+HNO=$C_6H_5NO$+H• | $3.79 \times 10^{9}$ | 1.19 | 95400 | Choi05 [68] | (25.) |
| $C_6H_5NO_2$ =$C_6H_5$•+$NO_2$ | $1.52 \times 10^{17}$ | 0.0 | 73717 | Xu05 [74] | (26.) |
| $C_6H_5NO$=$C_6H_5$•+NO | $1.52 \times 10^{17}$ | 0.0 | 55200 | Tseng04 [75] | (27.) |
| **Reactions of phenyl peroxy radicals** | | | | | |
| $C_6H_5O_2$•+NO=$C_6H_5O$•+$NO_2$ | $4.7 \times 10^{12}$ | 0.0 | -358.0 | Estimated[j] | (28.) |
| **Reactions of toluene and benzyl radicals** | | | | | |
| benzyl+$HNO_2$=toluene+$NO_2$ | $8.14 \times 10^{4}$ | 1.87 | 4838 | Estimated[k] | (29.) |
| benzyl+HONO=toluene+$NO_2$ | $8.14 \times 10^{4}$ | 1.87 | 5504 | Estimated[l] | (30.) |
| benzyl+HNO=toluene+NO | $1.47 \times 10^{10}$ | 0.76 | 349.0 | Estimated[m] | (31.) |
| •$C_6H_4CH_3$+HNO=toluene+NO | $3.78 \times 10^{5}$ | 2.28 | 456 | Estimated[n] | (32.) |
| benzyl+$NO_2$=$C6H5CH2O$• + NO | $1.36 \times 10^{12}$ | 0.0 | 0.0 | Estimated[o] | (33.) |
| $C_6H_5CH_2NO_2$=benzyl+$NO_2$ (high pressure) | $1.8 \times 10^{17}$ | 0.0 | 58500 | Estimated[p] | (34.) |
| (low pressure) | $1.3 \times 10^{18}$ | 0.0 | 42000 | | |
| **Reactions of benzylperoxy radicals** | | | | | |
| $C_6H_5CH_2OO\bullet + NO = NO_2 + C_6H_5CH_2O\bullet$ | $4.70 \times 10^{12}$ | 0.0 | -358 | Estimated[j] | (35.) |
| **Reactions of benzylalkoxy radicals** | | | | | |
| $C_6H_5CH_2O\bullet + NO = C_6H_5CHO + HNO$ | $7.6 \times 10^{13}$ | -0.76 | 0.0 | Estimated[q] | (36.) |
| $C_6H_5CH_2O\bullet + NO_2 = C_6H_5CHO + HONO$ | $4.0 \times 10^{12}$ | 0.0 | 2285 | Estimated[r] | (37.) |
| $C_6H_5CH_2O\bullet + HNO = C_6H_5CH_2OH + NO$ | $3.16 \times 10^{13}$ | 0.0 | 0.0 | Estimated[s] | (38.) |
| **Reactions of benzaldehyde radicals** | | | | | |
| $C_6H_5CHO + NO_2 = C_6H_5CO + HONO$ | $8.35 \times 10^{-10}$ | 6.68 | 8300 | Estimated[t] | (39.) |
| $C_6H_5CHO + NO = C_6H_5CO + HNO$ | $1.02 \times 10^{13}$ | 0.0 | 40670 | Estimated[u] | (40.) |
| **Reactions of $HOC_6H_4CH_2\bullet$ radicals** | | | | | |
| $HOC_6H_4CH_2\bullet + HNO = HOC_6H_4CH_3 + NO$ | $1.47 \times 10^{11}$ | 0.76 | 349.0 | Estimated[v] | (41.) |
| $HOC_6H_4CH_2\bullet + NO_2 = HOC_6H_4CH_2O\bullet + NO$ | $1.36 \times 10^{12}$ | 0.0 | 0.0 | Estimated[v] | (42.) |
| $HOC_6H_4CH_2\bullet + HONO = HOC_6H_4CH_3 + NO_2$ | $8.1 \times 10^4$ | 1.87 | 5504 | Estimated[w] | (43.) |
| **Reactions of $C_6H_5CH_2NO_2$** | | | | | |
| $C_6H_5CH_2NO_2 + \bullet OH = C_6H_5CHO + NO + H_2O$ | $3.0 \times 10^6$ | 2.0 | 2000 | Estimated[x] | (44.) |
| $C_6H_5CH_2NO_2 + \bullet O\bullet = C_6H_5CHO + NO + \bullet OH$ | $1.51 \times 10^{13}$ | 0.0 | 5354 | Estimated[x] | (45.) |
| $C_6H_5CH_2NO_2 + H\bullet = C_6H_5CHO + NO + H_2$ | $4.67 \times 10^{12}$ | 0.0 | 3732 | Estimated[x] | (46.) |
| $C_6H_5CH_2NO_2 + \bullet CH_3 = C_6H_5CHO + NO + CH_4$ | $7.08 \times 10^{11}$ | 0.0 | 11140 | Estimated[x] | (47.) |
| $C_6H_5CH_2NO_2 + \bullet CH_3 = HCHO + NO + toluene$ | $7.08 \times 10^{11}$ | 0.0 | 11140 | Estimated[x] | (48.) |
| **Reactions of $C_6H_5NO_2$** | | | | | |
| $C_6H_5NO_2 = C_6H_5O\bullet + NO$ | $7.12 \times 10^{13}$ | 0.0 | 62590 | Xu05 [74] | (49.) |
| $C_6H_5NO + NO_2 = C_6H_5NO_2 + NO$ | $9.62 \times 10^{10}$ | 0.0 | 12928 | Park02[76] | (50.) |

___

[a]: The rate constant was taken as 10 times the value proposed by Glänzer and Troe [77].

[b]: The rate constant was taken as 3.6 times the value proposed by Bruykov et al. [78] for the reaction between HNO and the $O_2$ radical.

[c]: The rate constant was taken as equal to that of reaction 2.

[d]: α is the number of primary H-atoms, β of secondary H-atoms and γ of tertiary H-atoms.

[e]: The rate constant was taken as equal to that of reaction 3.

[f]: The rate constant was taken similar to those proposed by Glarborg et al. [72] for the reaction $CH_3+NO_2=CH_3O\bullet+NO$.

[g]: The rate constant was taken as 1.8 times the value proposed by Atkinson et al. [61] for the reaction between $CH_3OO\bullet$ radicals and NO.

[h]: The •OOQOOH decomposition defined as : $C_nH_{(2n)}OOOOH + NO => NO_2 + OH + 2HCHO + C_{(n-2)}H_{(2n-4)}$.

[i]: The rate constant was taken as equal to that of reaction 6.

[j]: The rate constant was taken as equal to that of reaction 17.

[k]: The rate constant was taken as equal to 1/10[th] times the value proposed by Dean et al. [60] for the reactions of •$CH_3$ radicals with $HNO_2$.

[l]: The rate constant was equal to the value proposed by Dean et al. [60] for the reaction of HONO with •$CH_3$ radicals.

[m]: The rate constant was taken as equal to 1/10[th] times the value proposed by Choi et al. [68] for the reactions of •$CH_3$ radicals with HNO.

[n]: The rate constant was taken as equal to that of reaction 24.

[o]: The rate constant was taken as equal to that of reaction 2 divided by 10.

[p]: The rate constant was taken as equal to that of reaction 3.

[q]: The rate constant was taken as equal to that of reaction 4.

[r]: The rate constant was taken as the value proposed in GRIMECH 3.0 [59] for the reaction $CH_3O\bullet$ radicals.

[s]: The rate constant was taken as equal to that of reaction 11.

[t]: The rate constant was equal to 10 times the value taken for reaction 6.

[u]: The rate constant was equal to the value taken for reaction 5.

[v]: The rate constant was taken equal to that of the similar reaction for benzyl radicals.

[w]: The rate constant was taken as equal to 1/10[th] of the value proposed by Dean et al. [60] for the reactions of •$CH_3$ radicals with HONO.

[x]: The rate constant was equal to the value proposed in GRIMECH 3.0 [59] for the similar reaction of $CH_3NO_2$.

___

**TABLE II: SUMMARY OF EXPERIMENTAL CONDITIONS USED FOR AUTOIGNITION SIMULATIONS IN RAPID COMPRESSION MACHINES AND SHOCK TUBES**

| Author | Exp. setup | Fuel Reference | FUEL | | | Dilutant | Pressure | Temperature | Eq. Ratio |
| | | | n-heptane [mol %] | Isooctane [mol %] | toluene [mol %] | | [$10^5$ Pa] | [K] | [-] |
|---|---|---|---|---|---|---|---|---|---|
| Vanhove et al. [21] | RCM | Fuel_Van_1 | 100 | 0 | 0 | $CO_2$, $N_2$, Ar | 3.3 - 4.5 | 650 - 910 | 1.0 |
| | RCM | Fuel_Van_2 | 0 | 100 | 0 | | 12.4 - 15.8 | 660-880 | 1.0 |
| | RCM | Fuel_Van_3 | 50 | 0 | 50 | | 3.8 - 4.8 | 650 - 860 | 1.0 |
| | RCM | Fuel_Van_4 | 0 | 65 | 35 | | 12.0 - 14.6 | 670 - 850 | 1.0 |
| Callahan et al. [22] | RCM | Fuel_Cal_1 | 0 | 100 | 0 | $CO_2$, $N_2$, Ar | 11 - 17 | 650-910 | 1.0 |
| | RCM | Fuel_Cal_2 | 5 | 95 | 0 | | 11 – 17 | 650-910 | 1.0 |
| | RCM | Fuel_Cal_3 | 10 | 90 | 0 | | 11 - 17 | 650-910 | 1.0 |
| Tanaka et al. [30] | RCM | Fuel_Tan_1 | 26 | 0 | 74 | $N_2$ | 10 | 318 | 0.4 |
| | RCM | Fuel_Tan_2 | 21 | 5 | 74 | | 10 | 318 | 0.4 |
| | RCM | Fuel_Tan_3 | 16 | 10 | 74 | | 10 | 318 | 0.4 |
| Ciezki et al. [23] | ST | Fuel_Cie_1 | 100 | 0 | 0 | $CO_2$, $N_2$ | 13 | 650-1200 | 0.5-2 |
| Fieweger et al. [24] | ST | Fuel_Fiew_1 | 0 | 100 | 0 | $CO_2$, $N_2$ | 40 | 650-1200 | 1.0 |
| | ST | Fuel_Fiew_2 | 10 | 90 | 0 | | 40 | 650-1200 | 1.0 |
| | ST | Fuel_Fiew_3 | 20 | 80 | 0 | | 40 | 650-1200 | 1.0 |
| | ST | Fuel_Fiew_4 | 40 | 60 | 0 | | 40 | 650-1200 | 1.0 |
| | ST | Fuel_Fiew_5 | 100 | 0 | 0 | | 40 | 650-1200 | 1.0 |
| Gauthier et al. [22] | ST | Fuel_Gau_1 | 17 | 55 | 28 | $N_2$ | 860 - 1022 | 12 and 60 | 1.0 |
| | ST | Fuel_Gau_2 | 17 | 63 | 20 | | | 12 and 60 | 1.0 |

**TABLE III: SUMMARY OF EXPERIMENTAL CONDITIONS USED FOR SIMULATIONS IN HCCI ENGINES**

| Author | Exp. setup | Fuel Reference | n-C7 [mol%] | FUEL iso-C8 [mol%] | toluene [mol%] | Addit. NO [ppm] | Init. P. [$10^5$Pa] | Init. Temp. [K] | Eq. Ratio [ - ] | Engine Speed [rpm] | Compr. Ratio [-] |
|---|---|---|---|---|---|---|---|---|---|---|---|
| Andrae et al. [58] | HCCI | Fuel_Andr_1 (Run_1) | 6 | 94 | 0 | 0 | 1 | 393 | 0.29 | 900 | 16.7 |
| | HCCI | Fuel_Andr_1 (Run_2) | 6 | 94 | 0 | 0 | 2 | 313 | 0.25 | 900 | 16.7 |
| | HCCI | Fuel_Andr_2 (Run_2) | 16 | 84 | 0 | 0 | 2 | 313 | 0.25 | 900 | 16.7 |
| | HCCI | Fuel_Andr_3 (Run_1) | 25 | 0 | 75 | 0 | 1 | 393 | 0.29 | 900 | 16.7 |
| | HCCI | Fuel_Andr_3 (Run_2) | 25 | 0 | 75 | 0 | 2 | 313 | 0.25 | 900 | 16.7 |
| | HCCI | Fuel_Andr_4 (Run_2) | 35 | 0 | 65 | 0 | 2 | 313 | 0.25 | 900 | 16.7 |
| Dubreuil et al. [4] | HCCI | Fuel_Dub_1 | 100 | 0 | 0 | 0-500 | 1 | 348 | 0.3 | 1500 | 17 |
| | HCCI | Fuel_Dub_2 | 75 | 25 | 0 | 0-500 | 1 | 348 | 0.3 | 1500 | 17 |
| | HCCI | Fuel_Dub_3 | 80 | 0 | 20 | 0-500 | 1 | 348 | 0.3 | 1500 | 17 |
| Risberg et al. [41] | HCCI | Fuel_Ris_1 | 16 | 84 | 0 | 0-450 | 1 | 373 | 0.25 | 900 | 13.6 |
| | HCCI | Fuel_Ris_2 | 35 | 0 | 65 | 0.450 | 1 | 373 | 0.25 | 900 | 13.6 |

**TABLE IV: SUMMARY OF EXPERIMENTAL CONDITIONS USED FOR SIMULATIONS IN PERFECTLY STIRRED REACTORS (PSR)**

| Author | Exp. setup | Fuel Reference | n-C7 [mol%] | FUEL Iso-C8 [mol%] | toluene [mol%] | Additive NO [ppm] | Dilution | Pressure [$10^5$ Pa] | T [K] | Eq. Ratio [ - ] |
|---|---|---|---|---|---|---|---|---|---|---|
| Moréac [39] | PSR | Fuel_Morc_1 | 100 | 0 | 0 | 0, 50, 500 | 98 % N2 | 1 | 550-1100 | 1.0 |
| | PSR | Fuel_Morc_1 | 100 | 0 | 0 | 0, 50, 500 | 98 % N2 | 10 | 550-1100 | 1.0 |
| | PSR | Fuel_Morc_2 | 0 | 100 | 0 | 0, 50, 500 | 98 % N2 | 1 | 550-1100 | 1.0 |
| | PSR | Fuel_Morc_2 | 0 | 100 | 0 | 0, 50, 500 | 98 % N2 | 10 | 550-1100 | 1.0 |
| | PSR | Fuel_Morc_3 | 0 | 0 | 100 | 0, 50, 500 | 98 % N2 | 1 | 550-1100 | 1.0 |
| | PSR | Fuel_Morc_3 | 0 | 0 | 100 | 0, 50, 500 | 98 % N2 | 10 | 550-1100 | 1.0 |
| Dubreuil et al. [4] | PSR | Fuel_Dub_4 | 15 | 85 | 0 | 0, 50, 200 | 98 % N2 | 10 | 550-1100 | 0.2 |
| | PSR | Fuel_Dub_3 | 80 | 0 | 20 | 0, 50 | 9 8% N2 | 10 | 550-1100 | 0.2 |

# FIGURES

**Figure 1**

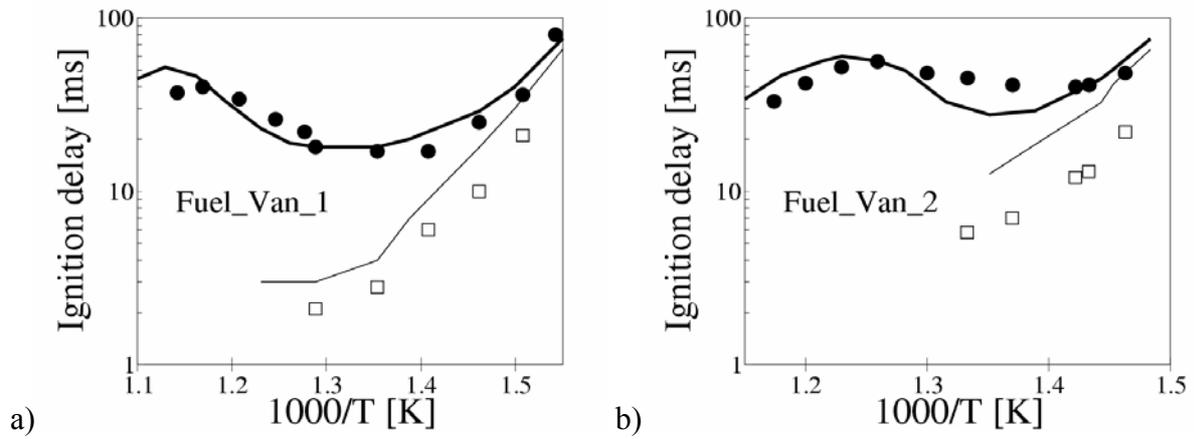

a)          b)

**Figure 2**

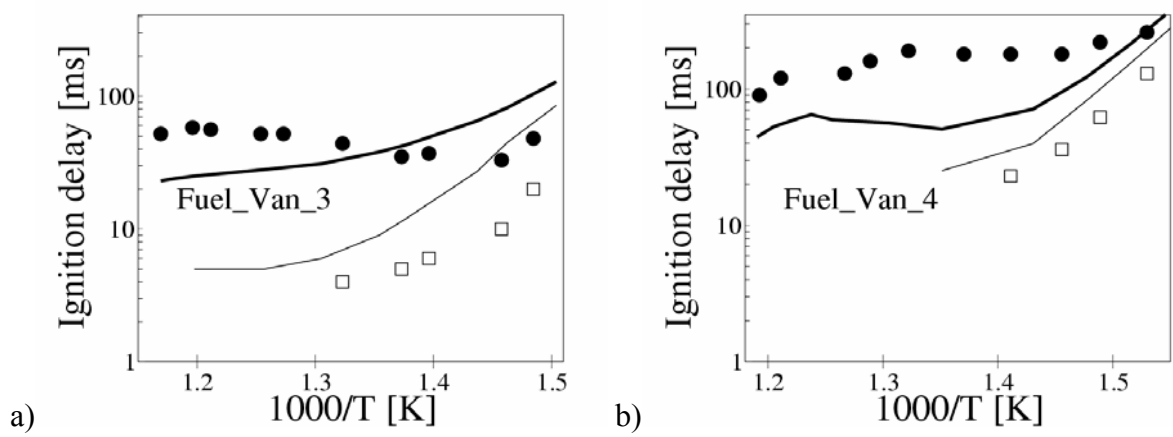

a)          b)

**Figure 3**

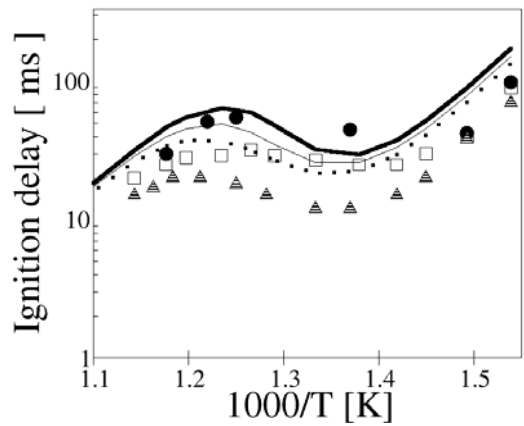

**Figure 4**

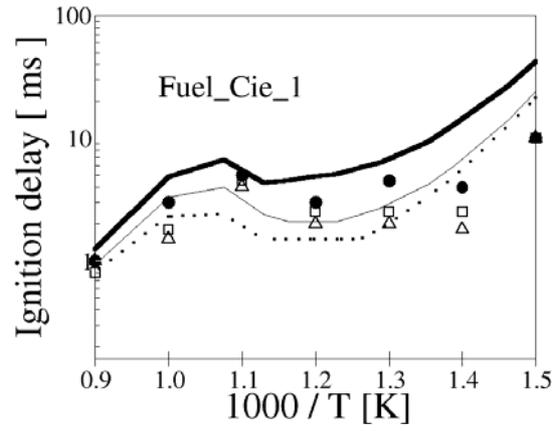

**Figure 5**

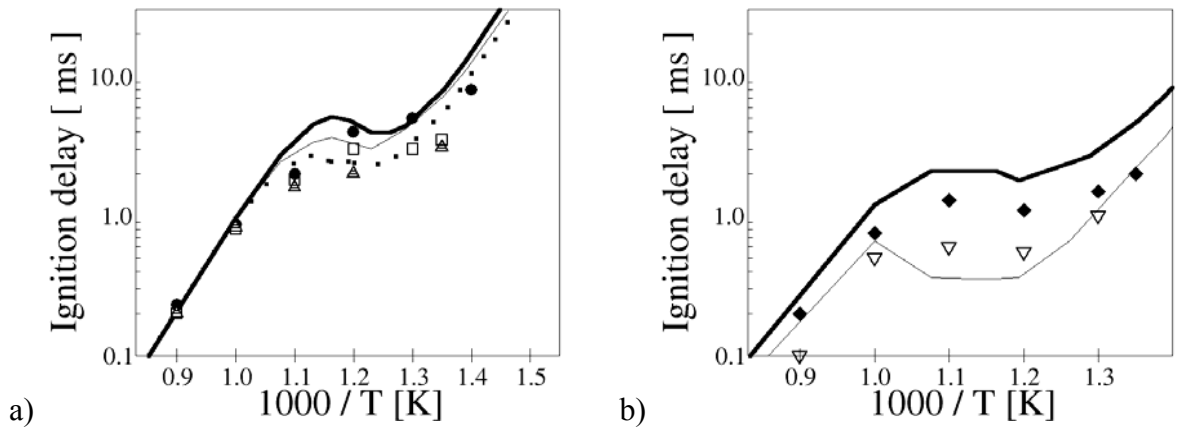

a) b)

**Figure 6**

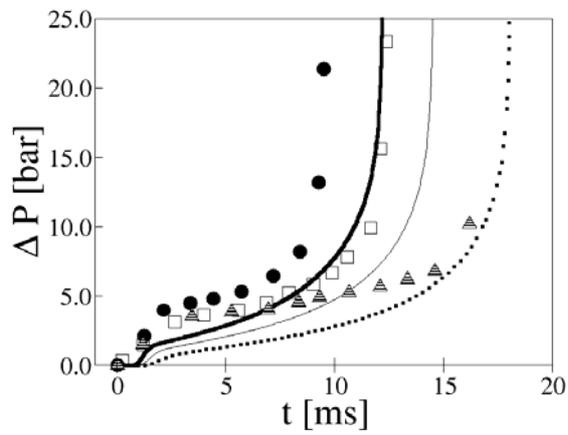

**Figure 7**

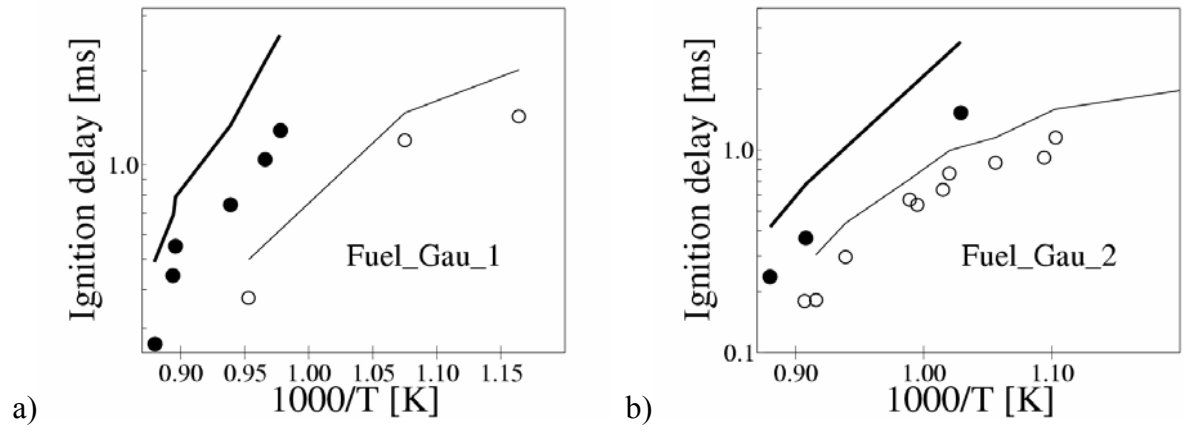

**Figure 8**

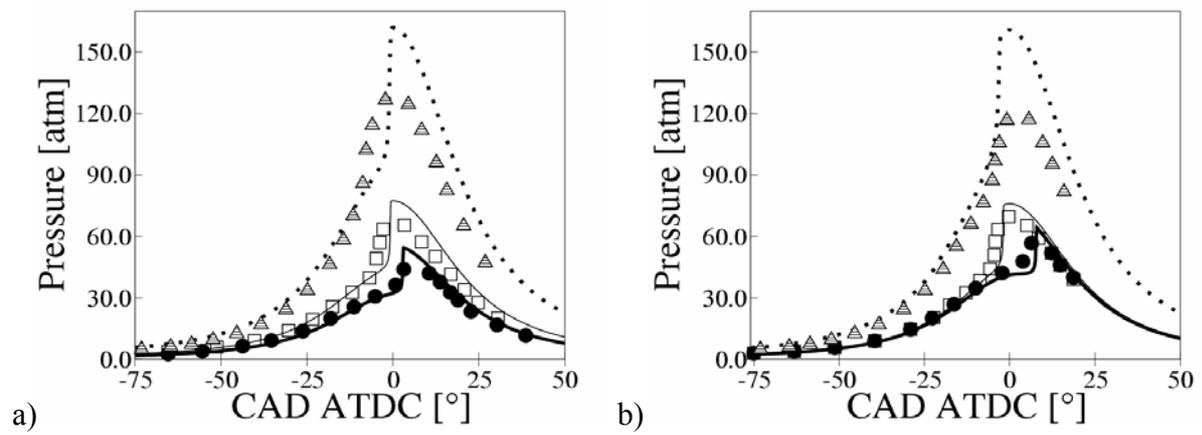

**Figure 9**

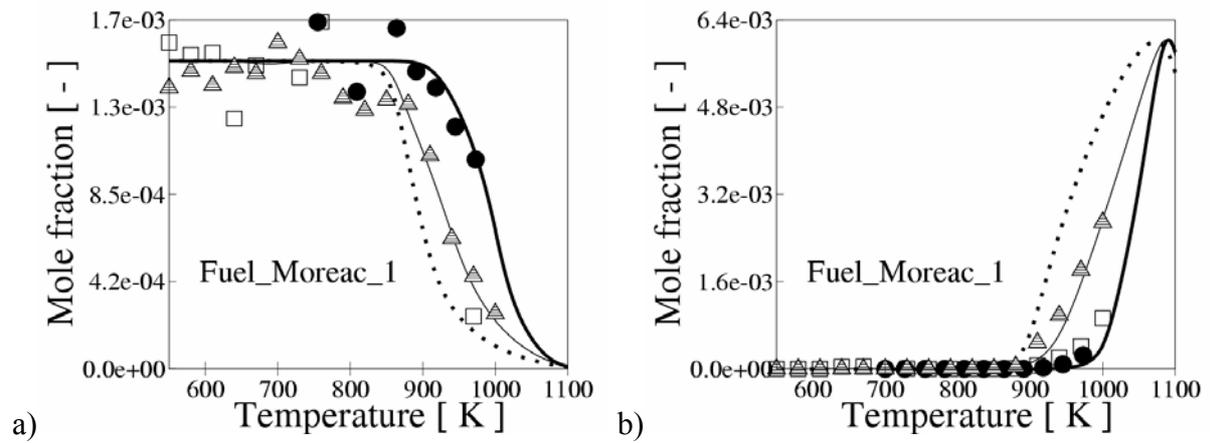

a) b)

**Figure 10**

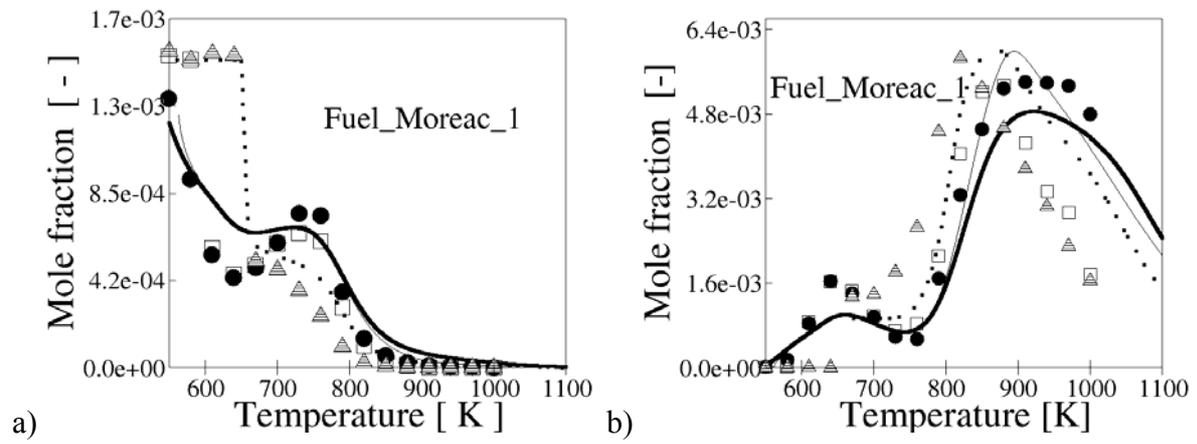

a) b)

**Figure 11**

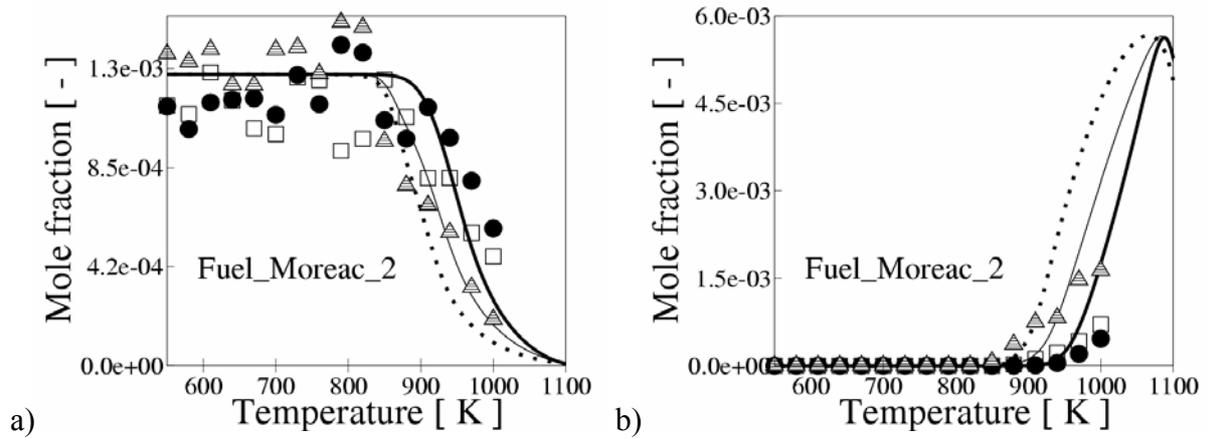

a) b)

**Figure 12**

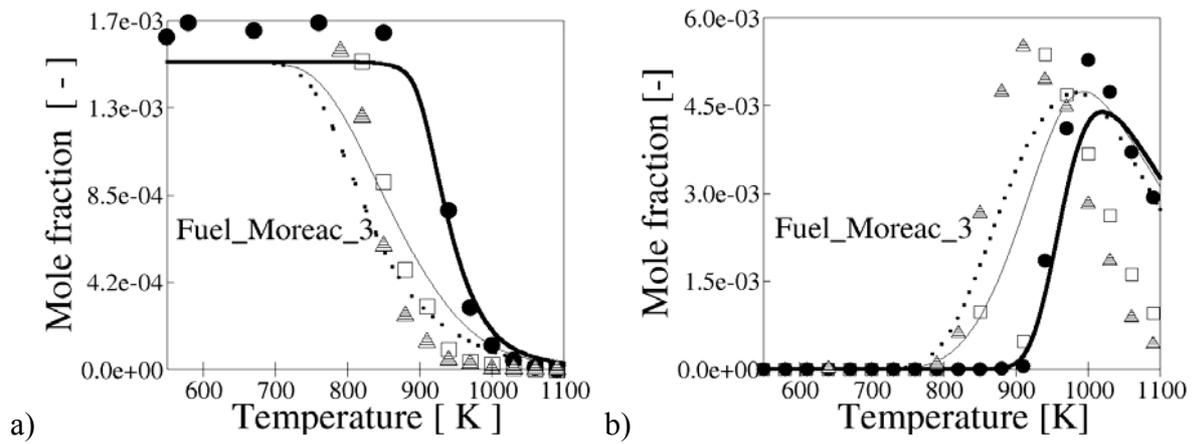

a) b)

**Figure 13**

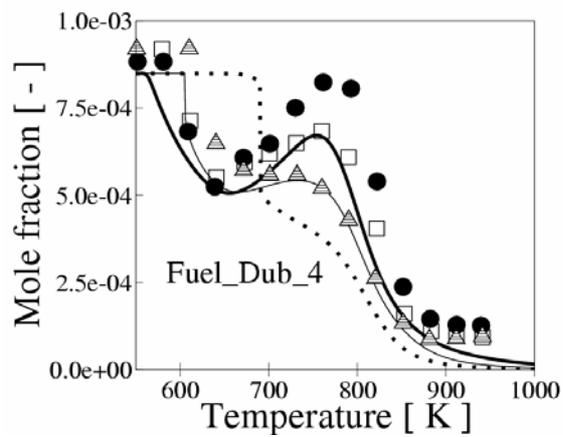

Figure 14

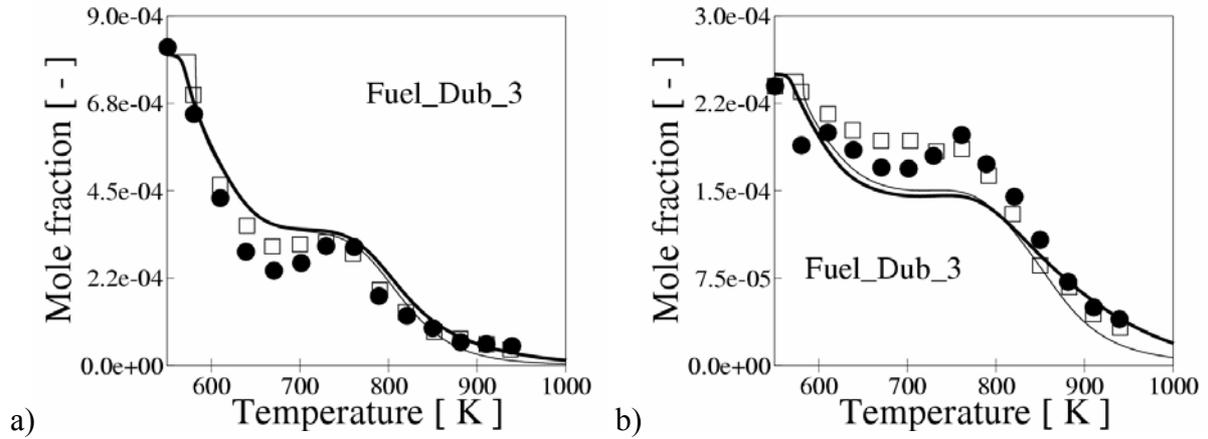

a) b)

Figure 15

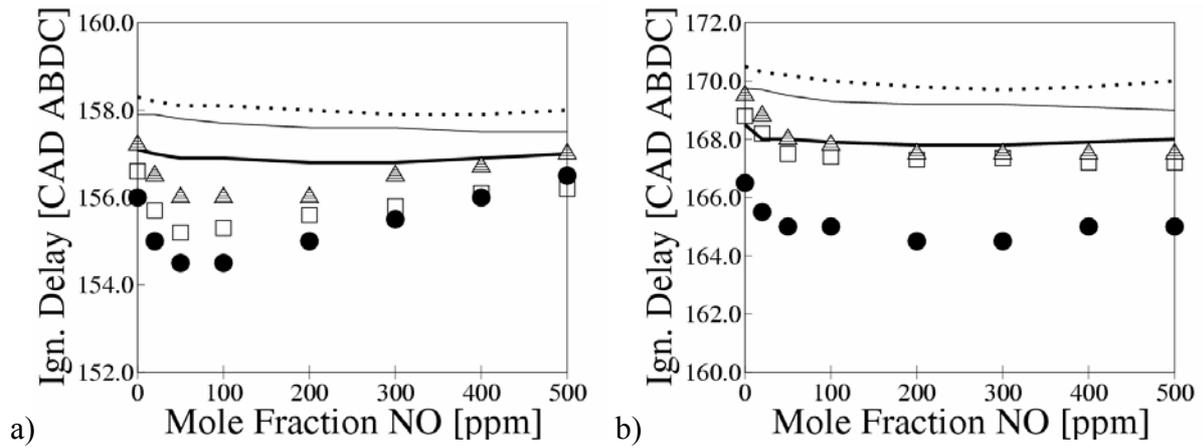

a) b)

Figure 16

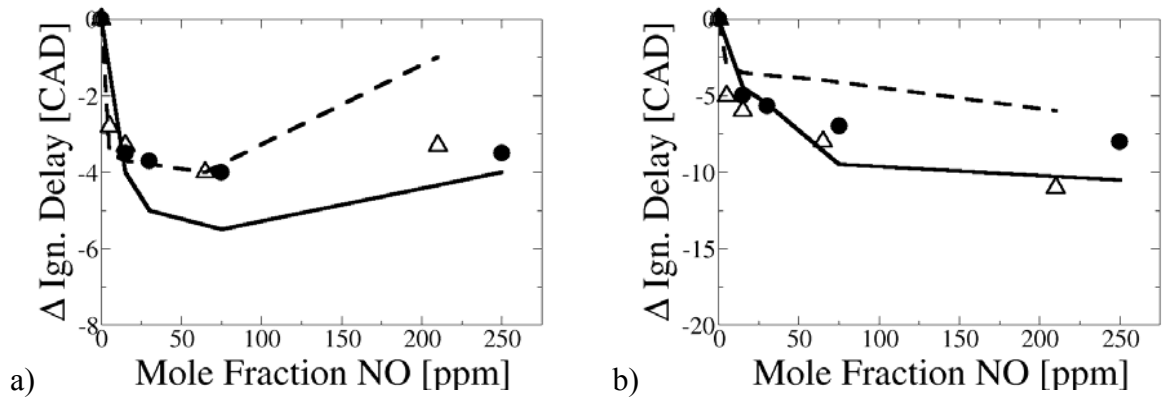

a) b)

**Figure 17**

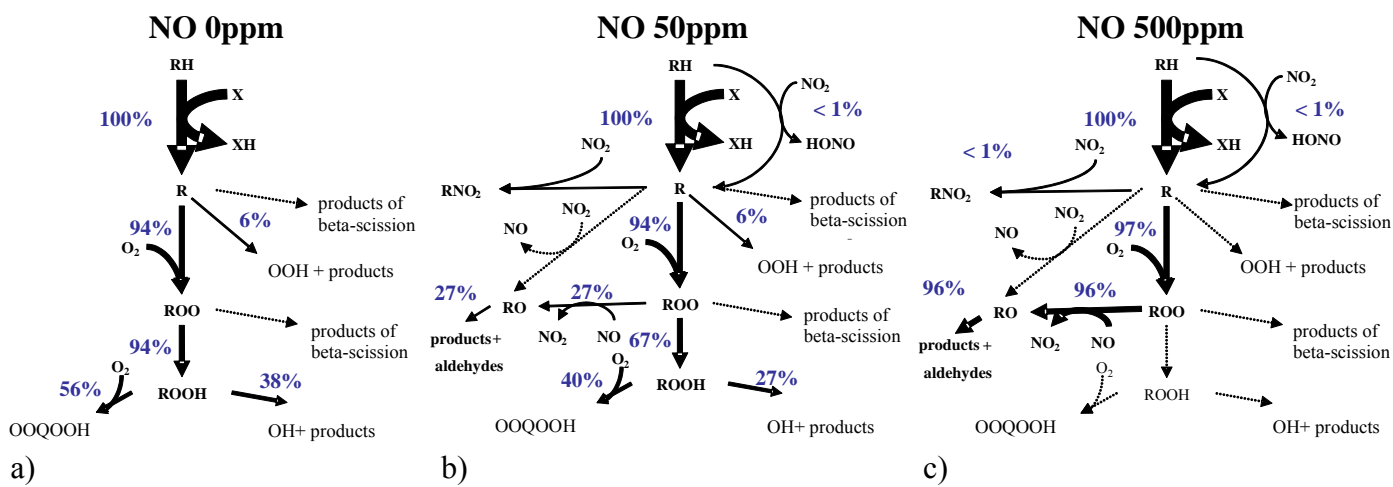

**Figure 18**

**Fuel: n-heptane, Pressure 1atm, T 665, 750, 900K**

NO 50 ppm          NO 500 ppm

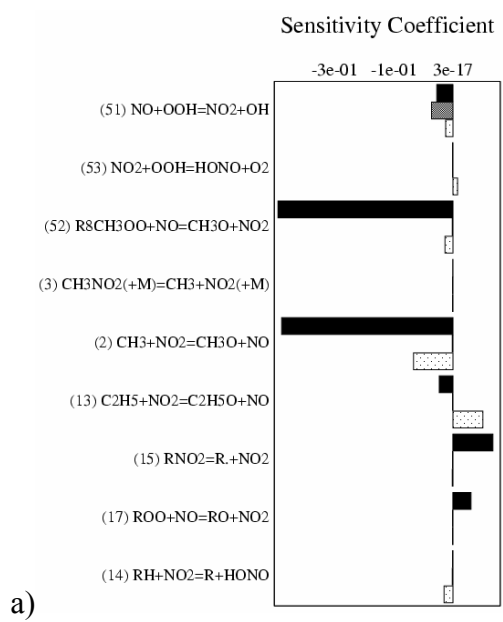
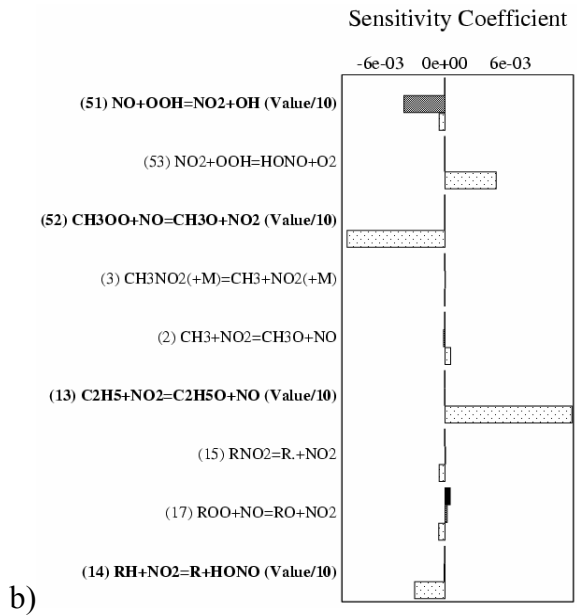

a)          b)

**Figure 19**

**Fuel: n-heptane, Pressure 10atm, T 665, 750, 900K**

**NO 50 ppm**   **NO 500 ppm**

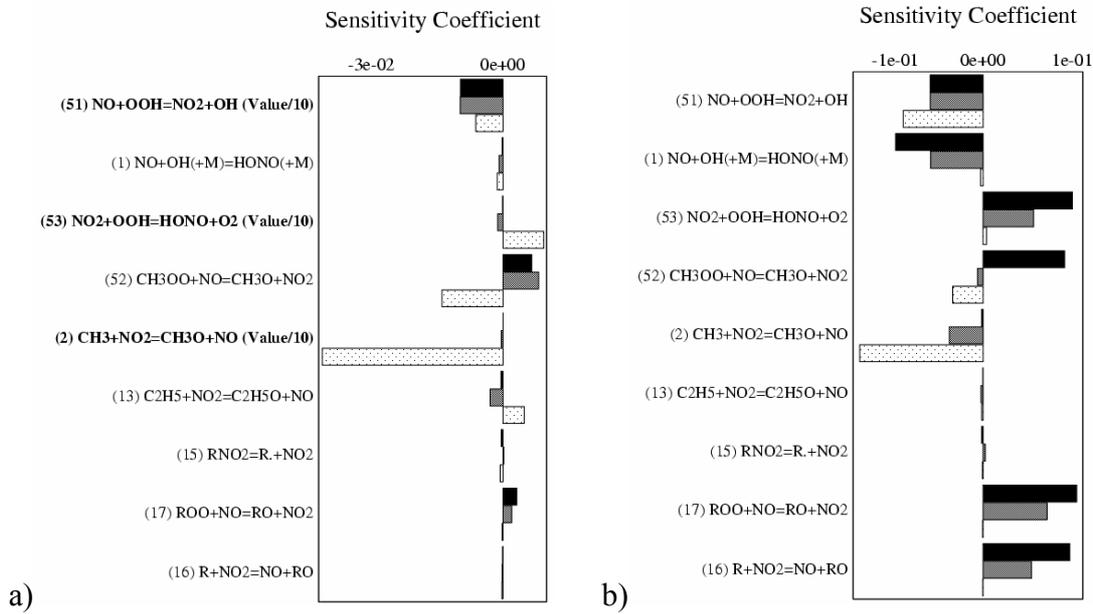

**Figure 20**

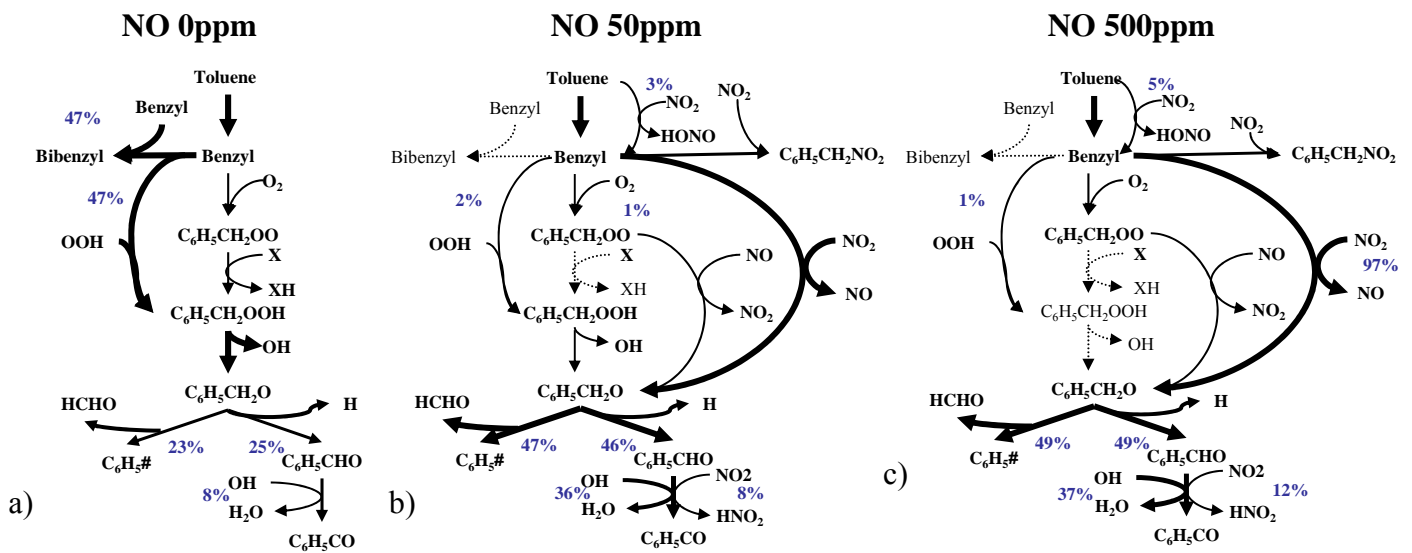

**Figure 21**

**Fuel: toluene, Pressure 10atm, T 750, 900, 1000K**

NO 50 ppm  NO 500 ppm

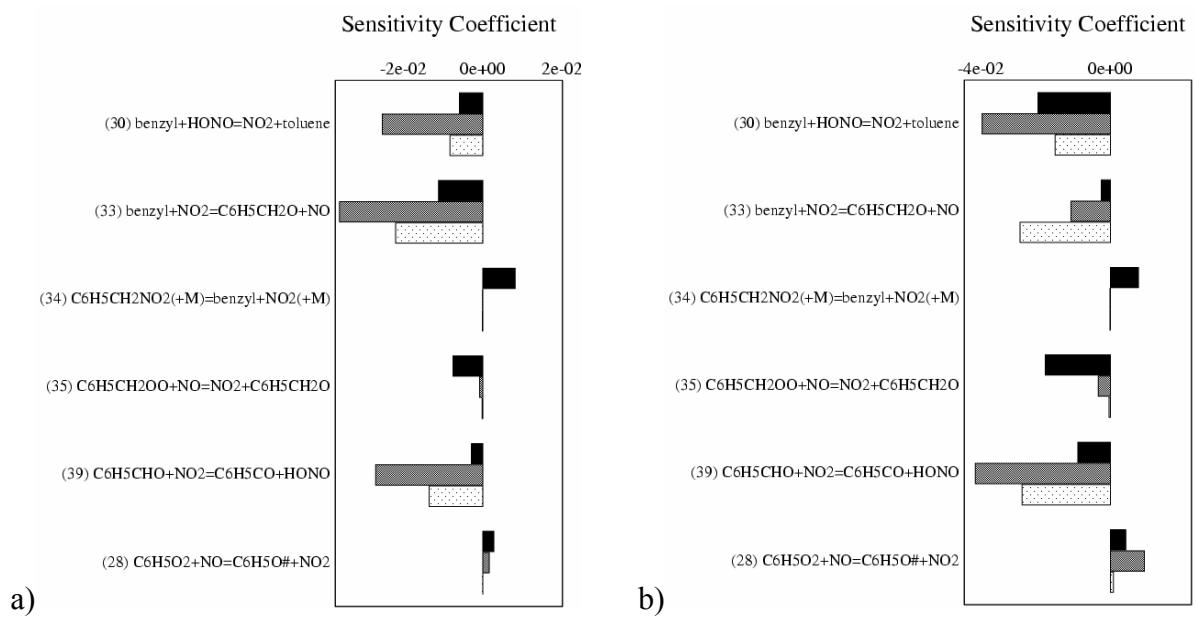

a)  b)

**FIGURE CAPTIONS**

**Figure 1**  Experimental (symbols) and simulated (lines) ignition delay times of cool flame- (white squares □ and thin full line -) and main flame- (black circles ● and thick full line ▬) ignition delays obtained in a **RCM** [21] for the stoichiometric oxidation of (a) **n-heptane** at pressures around **4 atm** and (b) **iso-octane** at pressures around **14 atm**.

**Figure 2**  Experimental (symbols) and simulated (lines) ignition delay times of cool flame (white squares □ and thin full line -) and main flame (black circles ● and thick full line ▬) ignition delays obtained in a **RCM** [21] for the stoichiometric oxidation of (a) **n-heptane/toluene** mixtures at pressures around **4 atm** and (b) **iso-octane/toluene** mixtures at pressures around **14 atm**.

**Figure 3**  Experimental (symbols) and simulated (lines) main flame ignition delay times obtained in a **RCM** [22] for the stoichiometric oxidation of PRF mixtures at pressures around **15 atm** with a **RON** of **100** (black circles ● and thick full line ▬), **95** (white squares □ and thin full line -) and **90** (grey shaded triangles ∆ and dotted lines ---).

**Figure 4**  Experimental (symbols) and simulated (lines) ignition delay times obtained in a **ST** [23] for **n-heptane** air mixtures at **13 atm**, for an **equivalence ratio** of **0.5** (black circles ● and thick full line ▬), **1.0** (white squares □ and thin full line -) and **2.0** (white triangles ∆ and dotted line ---).

**Figure 5**  Experimental (symbols) and simulated (lines) ignition delay times obtained in a **ST** [24] at around **40 atm** for the stoichiometric oxidation of **PRF** air mixtures with a **RON** of (a) **100** (black circles ● and thick full line ▬), **90** (white squares □ and thin full line -) **80** (grey triangles ∆ and dotted line ...) and (b) **60** (black diamonds ♦ and thick full line ━) and **0** (white triangles Δ and thin full line -).

**Figure 6** Experimental (symbols) and simulated (lines) pressure profiles obtained in a **RCM** [30] at a pressure of **10 atm** for the oxidation of **Fuel_Tan_1** (black circles ● and thick full line ━), **Fuel_Tan_2** (white squares □ and thin full line -) and **Fuel_Tan_3** (grey shaded triangles Δ and dotted line ---) at an equivalence ratio of 0.4.

**Figure 7** Experimental (symbols) and simulated (lines) main flame ignition delay times obtained in a **ST** [25] for the stoichiometric oxidation of (a) **Fuel_Gau_1** at intermediate pressures (**12-25 atm**) (black circles ● and thick line ━) and at high pressures (**45-60 atm**) (white circles ○ and thin line -) and (b) of **Fuel_Gau_2** at intermediate pressures (**12-25 atm**) (black circles ● and thick line ━) , at high pressures (**45-60 atm**) (small white circles ○ and thin line -).

**Figure 8** Experimental (symbols) and simulated (lines) pressure profiles obtained in a **test engine** [26] for (a) engine Run_1 with **Fuel_Andr_1** (black circles ● and thick full line ━), engine Run_1 with **Fuel_Andr_2** (white squares □ and thin lines -), engine Run_2 with **Fuel_Andr_2** (grey shaded triangles Δ and dotted lines ---) and (b) engine Run_1 with **Fuel_Andr_3** (black circles ● and thick full line ━), engine Run_1 with **Fuel_Andr_4** (white squares □ and thin full line -), engine Run_2 with **Fuel_Andr_4** (grey shaded triangles Δ and dotted line ---).

**Figure 9** Experimental (symbols) and simulated (lines) concentration profiles of (a) n-heptane mole fractions and (b) CO mole fractions obtained in a **JSR** [39] for the stoichiometric oxidation of 1500 ppm of **n-heptane** (**Fuel_Moreac_1**) at a pressure of **1 atm** for an addition of 0 ppm NO (black circles ● and thick full line ━), 50 ppm NO (white squares □ and thin full line -) and 500 ppm NO (grey shaded triangles Δ and dotted lines ---).

**Figure 10** Experimental (symbols) and simulated (lines) concentration profiles of (a) n-heptane mole fractions and (b) CO mole fractions obtained in a **JSR** [39] for the stoichiometric oxidation of 1500 ppm of **n-heptane** (**Fuel_Moreac_1**) at a pressure of **10 atm** for an addition of 0 ppm NO (black circles ● and thick full line ━), 50 ppm NO (white squares □ and thin full line -) and 500 ppm NO (grey shaded triangles Δ and dotted line ---).

**Figure 11** Experimental (symbols) and simulated (lines) concentration profiles of (a) iso-octane mole fractions and (b) CO mole fractions obtained in a **JSR** [39] for the stoichiometric oxidation of 1250 ppm of **iso-octane** (**Fuel_Moreac_2**) at a pressure of **1 atm** for an addition of 0 ppm NO (black circles ● and thick full line ━), 50 ppm NO (white squares □ and thin full line -) and 500 ppm NO (grey shaded triangles Δ and dotted line ---).

**Figure 12** Experimental (symbols) and simulated (lines) concentration profiles of (a) toluene mole fractions and (b) CO mole fractions obtained in a **JSR** [39] for the stoichiometric oxidation of 1500 ppm of **toluene** (**Fuel_Moreac_3**) at a pressure of **10 atm** for an addition of 0 ppm NO (black circles ● and thick full line ━), 50 ppm NO (white squares □ and thin full line -) and 500 ppm NO (grey shaded triangles Δ and dotted line ---).

**Figure 13** Experimental (symbols) and simulated (lines) concentration profiles of iso-octane mole fractions obtained in a **JSR** [4] for the lean oxidation (Φ=0.2) of **Fuel_Dub_4** at a pressure of **10 atm** for an addition of 0 ppm NO (black circles ● and thick full line ━), 50 ppm NO (white squares □ and thin full line -) and 200 ppm NO (grey shaded triangles Δ and dotted line ---).

**Figure 14** Experimental (symbols) and simulated (lines) concentration profiles of (a) n-heptane mole fractions and (b) toluene mole fractions obtained in a **JSR** [4] for the lean

oxidation (Φ=0.2) of **Fuel_Dub_3** at a pressure of **10 atm** for an addition of 0 ppm NO (black circles ● and thick full line ▬) and 50 ppm NO (white squares □ and thin full line -).

**Figure 15** Experimental (symbols) and simulated (lines) ignition delays of (a) cool flame and (b) main flame as function of added NO obtained in an **HCCI-engine** [4] for the oxidation of **Fuel_Dub_1** (black circles ● and thick full line ▬), **Fuel_Dub_2** (white squares ▬ and thin full line -) and **Fuel_Dub_3** (grey shaded triangles ∆ and dotted line ---).

**Figure 16** Experimental (symbols) and simulated (lines) sensitivity to NO addition of (a) cool flame and (b) main flame in [CAD] obtained in an **HCCI-engine** [4] for the oxidation of **Fuel_Ris_1** (black squares ● and thick full line ▬), **Fuel_Ris_2** (white triangles ∆ and dotted lines ---).

**Figure 17** Flow analysis for the stoichiometric oxidation of 1500 ppm of **n-heptane** in a **PSR** at a pressure of **10 atm** and a temperature of **665 K** with addition of (a) 0 ppm of NO, (b) 50 ppm of NO and (c) 500 ppm of NO.

**Figure 18** Sensitivity analysis for the stoichiometric oxidation of 1500 ppm of **n-heptane** in a **PSR** at a pressure of **1 atm** at a temperature of 665 K (black bars), 750 K (dark grey bars) and 900 K (grey bars) for (a), an initial concentration of NO of 50 ppm and (b) an initial concentration of NO of 500 ppm.

**Figure 19** Sensitivity analysis for the stoichiometric oxidation of 1500 ppm of **n-heptane** in a PSR at a pressure of **10 atm** at a temperature of 665 K (black bars), 750 K (dark grey bars) and 900 K (grey bars) for (a), an initial concentration of NO of 50 ppm and (b) an initial concentration of NO of 500 ppm.

**Figure 20** Flow analysis for the stoichiometric oxidation of 1500 ppm of **toluene** in a PSR at a pressure of **10 atm** and a temperature of 900 K with addition of (a) 0 ppm of NO, (b)

50 ppm of NO and (c) 500 ppm of NO.

**Figure 21** Sensitivity analysis for the stoichiometric oxidation of 1500 ppm of **toluene** in a PSR at a pressure of **10 atm** at a temperature of 750 K (black bars), 900 K (dark grey bars) and 1000 K (grey bars) for (a), an initial concentration of NO of 50 ppm and (b) an initial concentration of NO of 500 ppm.